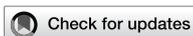

# A semi-Lagrangian method for the direct numerical simulation of crystallization and precipitation at the pore scale

Sarah Perez[1,2], Jean-Matthieu Etancelin[1] and Philippe Poncet[1]*

[1]Universite de Pau et des Pays de l'Adour, Centre National de la Recherche Scientifique (CNRS), Laboratoire de Mathématiques et de leurs Applications (LMAP, UMR5142), Pau, France, [2]The Lyell Centre, Heriot-Watt University, Edinburgh, United Kingdom

This article introduces a new efficient particle method for the numerical simulation of crystallization and precipitation at the pore scale of real rock geometries extracted by X-Ray tomography. It is based on the coupling between superficial velocity models of porous media, Lagrangian description of chemistry using Transition-State-Theory, involving underlying grids. Its ability to successfully compute dissolution process has been established in the past, and is presently generalized to precipitation and crystallization by means of adsorption modeling. Numerical simulations of mineral $CO_2$ trapping are provided, showing evidence of clogging/non-clogging regimes, and one of the main results is the introduction of a new non-dimensional number needed for this characterization.

KEYWORDS
digital rock physics (DRP), crystallization, mineral trapping, $CO_2$ storage, precipitation, clogging, Lagrangian methods, superficial velocity

# 1 Introduction

Studying reactive flows in porous media is essential to manage the geochemical effects arising from $CO_2$ capture and storage in natural underground reservoirs. While long-term predictions are commonly modeled at the field scale (Class et al., 2009), pore-scale approaches meanwhile provide insights into local geochemical interactions between the injected $CO_2$ and the aquifer structure (Payton et al., 2022). Through mathematical homogenization of the sub-micrometer porous medium and appropriate modeling, one can simulate the reactive processes that occur at the pore scale and predict their impact on the macro-scale properties, as developed in Allaire and Hutridurga (2012); Allaire et al. (2010). Geochemical processes are critical components for understanding the mineral trapping mechanisms and local evolving interfaces within the porous environment, and investigating the impact of such reactive processes provides insights into reservoir safety submitted to chemical interactions that may compromise the aquifer structure. Pore-scale modeling of reactive flow hence appears as a complementary mean to field scale studies wherein homogenization theory bridges the gap between these scales.

In this context, several geochemical mechanisms play a critical role in the $CO_2$ sequestration process and mainly involve precipitation, crystallization, and dissolution phenomena. On one side, carbonate precipitation and crystallization ensure efficient capture of the injected $CO_2$ in the form of minerals such as calcite, aragonite, or dolomite: this is referred to as mineral trapping, which informs about the storage capacities of the reservoir. These processes significantly impact the flow within the porous media at the pore scale, leading to restructuring of the flow path and morphological changes that alter, among other, the pore size distribution and the roughness of the interface due to partial or complete clogging of pore throats. Such alterations at the micro-scale subsequently alter the estimation of the macro-scale properties, namely, the porosity and permeability, and thereby require investigations to ensure wise management of the underground reservoir structures. On the other side, the reverse chemical process can also occur, resulting in carbonate mineral dissolution due to an acidification of the aqueous solution. This may compromise not only the efficiency of the trapping mechanisms, leading to an increase of both porosity and permeability, but also the integrity of the reservoir cap rock, and is, therefore, of great interest to prevent acute leakage issues. Consequently, one needs reliable estimations of the macro-properties changes due to these overall geochemical processes at the pore scale, to manage their impact on the reservoir scale modeling of $CO_2$ storage. This can be achieved through, first, efficient Direct Numerical Simulation (DNS) of reactive flows at the pore-scale and, subsequently, by embedding uncertainty concerns on the quantification of the petrophysical properties. In the present article, we address the first point with a focus on precipitation and crystallization modeling for $CO_2$ mineral storage into carbonate porous media.

Pore-scale investigation of reactive geochemical systems has garnered interest over the past decades based on imaging processes and laboratory experiments (Poonoosamy et al., 2023; Menke et al., 2015; Noiriel and Renard, 2022; Siena et al., 2021)), numerical simulations (Varzina et al., 2020; Patel et al., 2021; Payton et al., 2022; Soulaine et al., 2018), and a combination thereof (Molins et al., 2021; Noiriel and Soulaine, 2021). From this perspective, image-based DNS using microCT of a Representative Elementary Volume (REV) of a porous sample with efficient scientific computing and numerical method appears as a promising tool to query the impact of reactive processes on real rock geometries.

The present article focuses on the modeling aspects of $CO_2$ mineral trapping under the form of calcite crystal aggregates at the pore scale. Precipitation kinetics of calcite have been historically studied since the 1970s from the experimental and theoretical sides in Chou et al. (1989); Lasaga (1981); Plummer et al. (1978), and this has commonly established Transition State Theory (TST) as an efficient and straightforward way of predicting mineral reaction rate. Indeed, the deterministic TST is currently one of the most widely used models in reactive transport codes and DNS, detailed in Molins et al. (2012); Noiriel et al. (2021); Noiriel et al. (2016); Steefel et al. (2015). However, several doubts have risen in the research community about using such a deterministic approach for predicting crystal growth rates. In particular, comparison with experimentally determined growth rates has highlighted a wide range of discrepancies, querying the reliability of the TST model for overall crystallization processes, introduced in Hellevang et al. (2013); Pham et al. (2011). Meanwhile, probabilistic approaches, which find their origins in classical nucleation theory and the probabilistic nature of the precipitation and crystal growth mechanisms, have been developed in Masoudi et al. (2021); Nooraiepour et al. (2021a); Wolthers et al. (2012). These models make it possible to incorporate the effects of induction time characterizing the onset of the nucleation, the ionic affinities of the growing sites, and attachment frequencies of the ionic species involved in the reaction. Such attachment frequencies are, especially, significant for modeling surface adsorption and crystal aggregation that largely hinges on the surrounding porous structure in the sense that observable kinks and corners (above the voxel scale), for instance, are experimentally identified as preferential growing sites. However, such a geometrical dependency of the crystal aggregation is commonly neglected in most models, which makes it difficult to predict the spatial distribution of the new crystals.

Therefore, the main contribution of the present article is to build a novel numerical method dedicated to the precipitation and aggregation into crystal at the pore scale of microCT based geometry. Therefore, we formalize a two-step crystallization process wherein nuclei generation relies on a deterministic TST model before considering the probabilistic mineral aggregation—crystal growth—into the pore interface. The latter accounts for adsorption frequencies of the precipitate to the growth sites, which is weighted by a non-uniform probability of attachment rate depending on local mineral volume fraction. The current numerical method relies on a semi-Lagrangian approach, which handles a Lagrangian description of the chemistry with underlying grid methods for the hydrodynamic, based on the superficial velocity formalism introduced in the 1980s in Quintard and Whitaker (1988). The latter makes it possible to account for the involvement of the porous matrix in the overall flow process through a micro-continuum description of the medium. In this sense, one considers an intermediate state between the full resolution of each individual solid grain and the completely averaged continuum representation of the porous media at Darcy's scale. This establishes two-scale models that are widely used in hydrodynamics pore-scale modeling and $\mu$CT image-based DNS, as shown in Molins et al. (2021); Panga et al. (2005); Soulaine et al. (2017). The present semi-Lagrangian formalism has been successfully employed in the context of carbonate dissolution at the pore scale (Etancelin et al., 2020) and extensively benchmarked against state-of-the-art numerical alternatives, as detailed in Molins et al. (2021). The present work originality is to blend together the micro-continuum hydrodynamics and the two-step process involving precipitation with TST and aggregation with attachment probability. The result is an efficient numerical tool able to adress clogging studies in real geometries.

Moreover, this article investigates a case study involving a pure calcite sample in which alcalin solution saturated with calcium and dissolved $CO_2$ flow through, applying the numerical method developed here and allowing to compute whether this leads to material clogging or not. We show that the cohabitation of precipitation and aggregation define naturally two kinds of Damköhler numbers $Da_{II}$. This case study exhibits that as long as the precipitation Damköhler number is high enough, the crystallization Damköhler number drives whether there is clogging or not, for several Peclet numbers.

The present manuscript is structured as follows. The section 2 introduces reactive flow model at the pore-scale of rocks. First, the Darcy-Brinkman-Stokes model including to the Kozeny-Carman correlation terms in order to model highly heterogeneous medium at its pore scale, naturally meaningful for evolving fluid-solid interface defining the pores structure. Second, the transport-reaction-diffusion of chemical species. In Section 3, we detail our particle method for highly heterogeneous diffusion arising from Archie's law in the two-scale description of the medium. The application of such models to the precipitation and the crystallization for the $CO_2$ mineral trapping is described in Section 4. It includes the implementation and its HPC features on hybrid architectures, *i.e.*, coupling CPU and GPU devices, for the present application, detailed in Section 4.4. The related numerical results are discussed in Section 5 in terms of clogging or non-clogging regimes of crystallization.

## 2 Models in reactive microfluidics

The present section focuses on the modeling of reactive hydrodynamics in the context of $CO_2$ mineral storage and presents the mathematical model used to simulate reactive processes at the pore scale. We first introduce the so-called Darcy-Brinkman-Stokes formulation for microfluidic flows based on superficial velocity formalism. We subsequently incorporate transport-reaction-diffusion equations modeling the geochemical interactions between the different species involved. Finally, we present an alternative formulation in velocity-vorticity for the hydrodynamics equation, which ensures the fluid incompressibility condition.

### 2.1 Darcy-Brinkman-Stokes: a superficial velocity formalism at the pore scale

We introduce a spatial domain $\Omega \subset \mathbb{R}^n$, $n = 1, 2, 3$ which corresponds to the porous medium described at its pore scale. This sample description involves a pure fluid region $\Omega_F$, also called void-space and assumed to be a smooth connected open set, and a surrounding solid matrix $\Omega_S$ itself considered as a porous region. This region is seen as complementing the full domain $\Omega$, which in practice represents the computational box of the numerical simulations such that $\Omega_F = \Omega \setminus \overline{\Omega}_S$, and the internal fluid/solid interface is denoted $\Sigma$. We denote the computational domain boundary by $\partial\Omega$ and use $\Gamma_F = \partial\Omega \cap \Omega_F$ and $\Gamma_S = \partial\Omega \cap \Omega_S$ to refer to the fluid and solid parts of the computational domain boundary, respectively, such that $\partial\Omega = \Gamma_F \cup \Gamma_S$ (see Figure 1 for instance).

The boundary conditions at the inlet and outlet faces, typically for a cubic computational domain $\Omega = ]0, l[^3$ but not exclusively, either impose a prescribed flow rate $\overline{u}$ on the velocity or satisfy periodic boundary conditions for a prescribed driving force $f$. The boundary conditions on the other lateral faces are systematically periodic since rock samples are commonly constrained in an impermeable solid casing when $\mu$CT experiments are conducted. This ensures a consistent numerical representation of the sample compared to the experiments. This also guarantees $\mathcal{C}^\infty$ regularity on the boundary even if the domain exhibit corners, since the problem can be formalized by considering the equivalence relationship with the quotient space $\Omega \equiv \mathcal{Q}/G$ where $\mathcal{Q} = \mathbb{R}^2 \times ]0, l[$ and $G = l\mathbb{Z}^2 \times \{0\}$ (*e.g.*, see Sanchez et al. (2019) for detailed configurations of acceptable domains).

From the $\mu$CT images, we can also characterize the static pore-space structure, corresponding to the sample's initial state before any geochemical interactions. we denote by $\varepsilon = \varepsilon_f = 1 - \varepsilon_s$ the micro-porosity field defined on $\Omega$, given $\varepsilon_f$ and $\varepsilon_s$ respectively the volume fractions of void and solid according to usual notations from Soulaine et al. (2017). This defines a micro continuum description of the porous medium such that $\varepsilon = 1$ in the pure fluid region $\Omega_F$ and takes a small value in the surrounding matrix $\Omega_S$. In fact, the local micro-porosity $\varepsilon$ is assumed to have a strictly positive lower bound $\varepsilon(x, t) \geq \varepsilon_0 > 0$ for all $(x, t)$ in the spatiotemporal domain $\Omega \times [0, T_f]$ for a final real-time $T_f > 0$ in the reactive process. This lower bound $\varepsilon_0$ characterizes the residual porosity of the porous matrix, potentially unresolved due to X-ray $\mu$CT imaging limitations (as discussed in Perez et al. (2022), see also Figure 1). In practice, we assume throughout this work $\varepsilon_0 = 5\%$, but this value can range from 1% to 10%.

Such a two-scale description of the local heterogeneities in the carbonate rocks is appropriate to simulate the pore-scale physics and establish the governing flow and transport equations in each distinct region. Indeed, although the hydrodynamic of a viscous flow in a pure fluid region is commonly quantified through the Navier-Stokes equation, we can formulate the problem on the whole domain $\Omega$ based on the two-scale micro continuum description of the medium. We, therefore, consider the model on the superficial velocity $u$ introduced and derived rigorously by Quintard and Whitaker in the late 80 s (Quintard and Whitaker, 1988) and commonly used until nowadays (Molins et al., 2021; Soulaine et al., 2017; Wood et al., 2007):

$$\varepsilon^{-1}\frac{\partial \rho u}{\partial t} + \varepsilon^{-1}\text{div}\left(\varepsilon^{-1}\rho u \otimes u\right) - \varepsilon^{-1}\text{div}\left(2\mu D(u)\right) + \mu^* K_\varepsilon^{-1} u = f - \nabla p \quad (1)$$

along with the divergence-free condition $\text{div } u := \nabla \cdot u = 0$. It is noticeable that this incompressibility condition should be changed when considering evolving porous structures to account for density variations, especially in the context of fast dissolution or nucleation (Soulaine et al., 2018). Indeed, this only depicts that crystal nucleation within a liquid volume, for instance, drastically increases the density and induces divergence effects in its neighborhood. Nevertheless, the divergence-free condition can be assumed to remain if the characteristic time of fluid/solid interface changes is way larger than the hydrodynamics time scale (Soulaine et al., 2017), which is the case for our study. In Equation 1, the notation $D(u)$ refers the shear-rate tensor $D(u) = (\nabla u + \nabla u^T)/2$, $\mu$ is the dynamic viscosity, $p$ is the volumic pressure, $f$ the volumic driving force and $\rho$ the fluid density. The related viscosity $\mu^*$ usually coincides with the fluid viscosity $\mu$ but may be different in order to account for viscous deviations.

The quantities $\rho$, $\mu$, $\mu^*$ and $f$ are assumed to be constant. The quantities $\varepsilon$, $\rho$, $\mu$ and $p$ are scalar fields; $u$, its rotational $\omega = \nabla \times u$, $f$ and $\nabla p$ are vector fields, while $D(u)$ and $K_\varepsilon$ are matrices.

The permeability $K_\varepsilon$ refers to the micro-scale permeability and depends on the local micro-porosity field $\varepsilon$. This permeability of the micro-porous domain can be modeled by the empirical Kozeny-Carman relationship, historically introduced in Kozeny (1927) and Carman (1937), generalized in Quintard and Whitaker (1993) and now used meaningfully at the micrometer subscale (see Soulaine et al. (2017); Molins and Knabner (2019) and their next

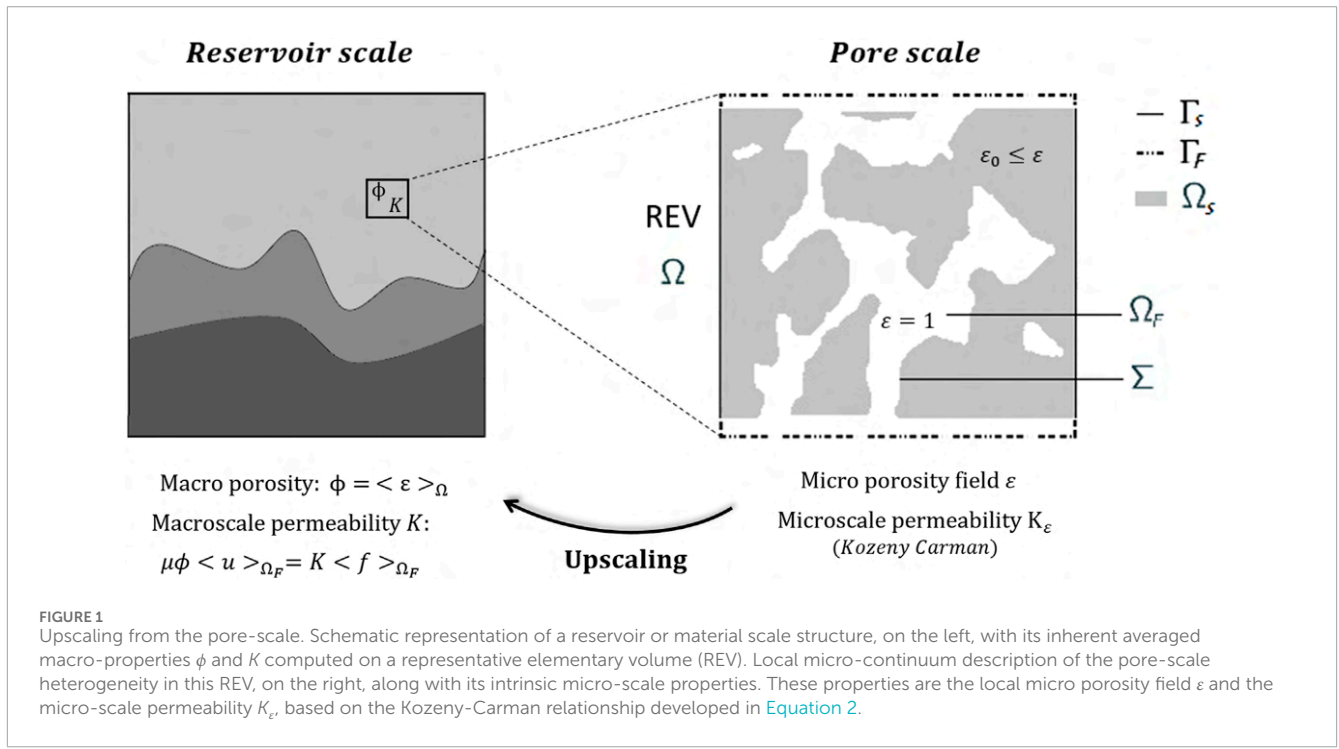

FIGURE 1
Upscaling from the pore-scale. Schematic representation of a reservoir or material scale structure, on the left, with its inherent averaged macro-properties $\phi$ and $K$ computed on a representative elementary volume (REV). Local micro-continuum description of the pore-scale heterogeneity in this REV, on the right, along with its intrinsic micro-scale properties. These properties are the local micro porosity field $\varepsilon$ and the micro-scale permeability $K_\varepsilon$, based on the Kozeny-Carman relationship developed in Equation 2.

publications). This relation is given by

$$K_\varepsilon^{-1} = \kappa_b^{-1} \frac{(1-\varepsilon)^2}{\varepsilon^3} \quad (2)$$

where $\kappa_b$ is the bulk permeability, which can be taken as a coarse estimation of the reference macro-scale permeability $\kappa_0$. For instance, Soulaine et al. (2017) estimated that four orders of magnitude below the permeability are required for $\kappa_b$ to ensure adherent boundary conditions at the pore interface. In this article, we consider both $K_\varepsilon$, $\kappa_b$ and $\kappa_0$ as scalars, meaning we restrict ourselves to the isotropic case although this formalism can be extended to anisotropic porous media. Despite the validation of this correlation law for porous media at the sub-millimeter scale and its applicability to dissolution (Molins et al., 2021), we assume that it is also valid for crystal growth at pore walls as a reverse mechanism of dissolution. The dissolution-crystallization (whether it is one or two steps process) is not reversible since the resulting crystal may not reach the same tortuosity than the initial material (hence a different permability), but the similarity in the material is sufficiently strong to work with this assumption. The superficial velocity formulation Equation 1 defines a two-scale model that can be solved on the overall domain $\Omega$ — using, for instance, penalization principles—and retrieves the usual Navier-Stokes equation in the pure fluid region $\Omega_F$ (since $K_\varepsilon^{-1} = 0$ for $\varepsilon = 1$). Depending on the flow regime hypothesis, one can also encounter simplified versions of Equation 1 wherein some terms can be neglected. In the context of pore-scale simulations, in particular, the inertial effects become negligible compared to viscous forces due to low Reynolds number, denoted Re. The latter is a characteristic dimensionless number defined as:

$$\text{Re} = \rho \bar{u} L/\mu,$$

where $\bar{u}$ and $L$ are respectively the characteristic velocity and length of the sample. The characteristic length $L$ can be related to average pore throat diameters and, therefore, we typically fall within the assumption Re $\ll$ 1 throughout this manuscript.

At low Reynolds numbers and for highly viscous Darcian flows, Equation 1 hence reduces to the following Darcy-Brinkman-Stokes (DBS) model:

$$-\text{div}\left(2\mu D(u)\right) + \mu \kappa_b^{-1} \frac{(1-\varepsilon)^2}{\varepsilon^2} u = \varepsilon(f - \nabla p), \quad \text{in} \quad \Omega \quad (3)$$

where $\mu^* = \mu$ for sake of readability. In the present work, we consider this DBS Equation 3, which is adequate in the flow regime hypothesis of low Reynolds number representative in pore-scale modeling. The DBS equation based on the superficial velocity is an efficient formalism to model the hydrodynamic in multi-scale porous media, and account for heterogeneous porosity levels.

## 2.2 Reactive flow model: general formulation

The DBS flow model Equation 3 needs to be complemented by transport-reaction-diffusion equations of the different species involved in the geochemical processes. These equations are derived from the mass balance of the chemical species (Soulaine et al., 2017), and can be written under the form:

$$\frac{\partial \varepsilon \widetilde{C}_k}{\partial t} + \text{div}\left(u\widetilde{C}_k\right) - \text{div}\left(\alpha_k(\varepsilon)\varepsilon\nabla\widetilde{C}_k\right) = \dot{m}_k/M_k, \quad (4)$$

where $\widetilde{C}_k = \rho_f \overline{\omega}_{f,k}/M_k$ is a concentration per unit of fluid with $M_k$ the molar mass and $\dot{m}_k$ the rate of mass transfer for the $k^{\text{th}}$ species. We follow here the notations introduced by Quintard and Whitaker in Quintard and Whitaker (1988), and afterward commonly used, where $\rho_f$ is the fluid density and $\overline{\omega}_{f,k}$ is the mass fraction of the $k^{th}$ component averaged on the fluid phase. The term $\alpha_k(\varepsilon)$ is a space-variable effective diffusion coefficient and accounts for a reduced diffusion in the surrounding porous matrix due to the tortuosity effect, which is usually quantified using Archie's law, historically introduced in Archie (1942):

$$\alpha_k(\varepsilon) = D_{m,k}\varepsilon^\eta.$$

In this empirical relationship, $\eta$ refers to the tortuosity index and $D_{m,k}$ to the molecular diffusion of the $k^{th}$ considered species, quantified at milimeter scale in Wakao and Smith (1962), generalized and downscaled in Coindreau and Vignoles (2005) and Glover (2016). We finally introduce $C_k = \varepsilon \widetilde{C}_k$, so that the Equation 4 is written:

$$\frac{\partial C_k}{\partial t} + \text{div}\left(\varepsilon^{-1} u C_k\right) - \text{div}\left(D_{m,k}\varepsilon^{1+\eta}\nabla\left(\varepsilon^{-1}C_k\right)\right) = R_k(\mathbf{C}), \quad (5)$$

which is no more than a superficial modeling of the chemistry, that is to say $C_k$ is the amount of moles per unit of volume while $\widetilde{C}_k$ is the amount of moles per unit of fluid volume. The notation $R_k(\mathbf{C})$ refers to a function (without differential operators) that models the rate contribution of the chemical reactions for the $k^{th}$ component, where we denote by $\mathbf{C} \in \mathbb{R}^{N_s}$ the vector of the concentrations $C_k$ of all the $N_s$ chemical species. We distinguish $N_m$ mobile species, and $N_i$ immobile species such that $N_s = N_m + N_i$. The $k^{th}$ rate contribution $R_k(\mathbf{C})$ is, practically, the balance of kinetics of all reactions involving the $k^{th}$ species. The sign of individual reaction rates lies in the nature of the species $k$ considered, either positive for a chemical product or negative for a reactant.

The model Equation 5 is the formalism that we retain for the aqueous species in the liquid phase. In particular, this model highlights a superficial gradient operator denoted $\nabla^\varepsilon \coloneqq \varepsilon\nabla\varepsilon^{-1}$ involved in the heterogeneous diffusion arising from the Archie's law. One should notice that the superficial gradient can become highly sensitive at the mineral boundary, mainly due to jumps in the porosity levels on either side of the interface, and thus will require special considerations to adequately manage evolving medium under reactive processes.

Concerning the solid phase of concentration $C_{(s)}$ (e.g., the $k^{th}$ component in vector $\mathbf{C}$), which we assume contains only one chemical species of molar volume $v$ ($N_i = 1$), it is not subject to transport or diffusion, so that

$$\frac{\partial C_{(s)}}{\partial t} = R_k(\mathbf{C}). \quad (6)$$

This solid concentration is subsequently linked to the micro-porosity $\varepsilon$ by the relation $C_{(s)} = (1-\varepsilon)/v$, so one gets

$$\frac{\partial \varepsilon}{\partial t} = -v R_k(\mathbf{C}).$$

In the case of a typical reaction involving a unique solid $X_{(s)}$ of molar volume $v$, and two aqueous species Y and Z in the liquid phase, with their respective positive stoechiometric coefficients $\chi_i$ and following, for instance, the general chemical reaction:

$$\chi_1 X_{(s)} + \chi_2 Y \rightleftharpoons \chi_3 Z, \quad (7)$$

we define the vector of concentrations $\mathbf{C} \coloneqq (C_1, C_2, C_3)^T = \left([X_{(s)}], [Y], [Z]\right)^T \in \mathbb{R}^3$. Since there is only one reaction, one gets a unique kinetic balance written as $R_i(\mathbf{C}) = \pm \chi_i R(\mathbf{C})$, with $R(\mathbf{C})$ the kinetic rate. By default, we assume a positive sign for the solid species, so that one follows the convention $R(\mathbf{C}) < 0$ for the forward reaction corresponding to the solid $X_{(s)}$ dissolution, while $R(\mathbf{C}) > 0$ for the reverse reaction, e.g., precipitation and crystallization processes. The sign for the aqueous species subsequently depends on its interaction with the solid X(s): we get a positive sign for species Y, which is either consumed or produced in the same way as the solid, and a negative sign for species Z, which behaves oppositely. The reaction rate $R(\mathbf{C})$ can involve many concentrations, specific areas, chemical activities, equilibrium constants, etc. (see Section 4 thereafter for practical examples and further details).

Along with its boundary and initial conditions, the model for reaction (Equation 7) defines a set of partial differential equations modeling reactive flows at the pore scale:

$$\begin{cases} -\text{div}(2\mu D(u)) + \mu\kappa_b^{-1}\dfrac{(1-\varepsilon)^2}{\varepsilon^2}u = \varepsilon(f - \nabla p), & \text{in } \Omega\times ]0, T_f[ \\ \dfrac{\partial C_1}{\partial t} = \chi_1 \ R(\mathbf{C}), & \text{in } \Omega\times ]0, T_f[ \\ \dfrac{\partial C_2}{\partial t} + \text{div}(\varepsilon^{-1}uC_2) - \text{div}(D_{m,2}\varepsilon^{1+\eta}\nabla(\varepsilon^{-1}C_2)) = \chi_2 R(\mathbf{C}), & \text{in } \Omega\times ]0, T_f[ \\ \dfrac{\partial C_3}{\partial t} + \text{div}(\varepsilon^{-1}uC_3) - \text{div}(D_{m,3}\varepsilon^{1+\eta}\nabla(\varepsilon^{-1}C_3)) = -\chi_3 R(\mathbf{C}), & \text{in } \Omega\times ]0, T_f[ \\ \varepsilon = 1 - vC_1, & \text{in } \Omega\times ]0, T_f[ \\ + \text{ adequate boundary and initial conditions, along with div } u = 0 \end{cases}$$
(8)

which is strongly coupled, since $u$ and $\mathbf{C}$ depend on each other by means of the micro-porosity changes $\varepsilon$ during the chemical process. It is also possible to straightforwardly substitute $C_1$ with $\varepsilon$ in this system (Equation 8). Finally, one can notice that the reactive system is valid on the whole domain $\Omega$, whether the local state is fluid or not. In the pure fluid region, this system indeed converges toward a Stokes hydrodynamic model coupled with a standard transport-diffusion equation. Mathematical modeling of reactive hydrodynamics at the pore-scale can be expressed under the general form of the PDE system (Equation 8) coupling DBS with transport-diffusion-reaction equations. It is noticeable that the micro-porosity $\varepsilon$ remains in the range $[\varepsilon_0, 1]$ which provides a well-posed Darcy-Brinkman-Stokes equation for the flow due to the expressions of Kozeny-Carman term and reaction formula.

This can be extended naturally to systems of reactions involving as many aqueous species in the liquid phase as needed, and involving potentially several solids: in this case the porosity is a linear combination of solid species. Most of the configurations studied in this article involve solid calcite—or calcium carbonate—whose concentration is denoted $C_{\text{CaCO}_{3(s)}}$ or $[\text{CaCO}_3]$, and whose molar volume is given by $v = 36.93 \times 10^{-3}$ L.mol$^{-1}$.

## 2.3 A velocity-vorticity formulation

Two distinct approaches are successfully used in the literature to solve numerically the DBS Equation 3, namely, the velocity-pressure or velocity-vorticity formulations (Hume and Poncet, 2021; Lamichhane, 2013; Molins et al., 2021; Angot, 2018). The latter, inherited from vortex methods (Chatelain et al., 2008; Cottet et al., 2000; El Ossmani and Poncet, 2010; Hejlesen et al., 2015) introduces the vorticity field $\omega$ which is intrinsically related to the fluid velocity $u$ through the relation:

$$\omega = \nabla \times u.$$

Several advantages arise when considering the velocity-vorticity formulation that regards the PDE unknowns $(u, \omega)$ and can be interpreted as describing the local spinning motions generated by the flow constraints. First of all, one can benefit from the velocity projection on divergence-free fields, and thereby analytically ensures the incompressibility condition div $u = 0$. Secondly, this formalism can be effectively coupled with splitting strategies that sequentially separate the resolution of distinct physical phenomena, such as convection and diffusion. Finally, this also makes it possible to eliminate the pressure unknown from the momentum equation by applying the curl operator on the DBS equation, which reads as follows:

$$-\mu \Delta \omega + \mu \kappa_b^{-1} \nabla \times \left( \frac{(1-\varepsilon)^2}{\varepsilon^2} u \right) = \nabla \varepsilon \times (f - \nabla p) \quad (9)$$

given the assumption $\nabla \times f = 0$. Developing the curl of the Kozeny-Carman term, one gets the following expression:

$$\nabla \times \left( \frac{(1-\varepsilon)^2}{\varepsilon^2} u \right) = \frac{(1-\varepsilon)^2}{\varepsilon^2} \omega + 2(\varepsilon - 1)\varepsilon^{-3} \nabla \varepsilon \times u$$

which, in practice, exhibits terms that become dominant compared to $\nabla \varepsilon \times (f - \nabla p)$. Consequently, the right-hand side in the vorticity formulation of the DBS Equation 9 is usually neglected (Etancelin et al., 2020; Molins et al., 2021).

Equation 9 is then supplemented with an equation that retrieves the velocity field from the related vorticity, and results in the relation:

$$-\Delta u = \nabla \times \omega \quad (10)$$

Based on the incompressibility condition. In practice, the previous Poisson Equation 10 is not straightforwardly considered, and one relies on an alternative using a stream function $\psi : \Omega \subset \mathbb{R}^3 \to \mathbb{R}^3$ (a vector potential) solution of:

$$\begin{cases} -\Delta \psi = \omega, & \text{in } \Omega \\ +\text{boundary conditions such that div } \psi = 0 \text{ on } \partial\Omega. \end{cases}$$

The condition div $\psi = 0$ on $\partial\Omega$ is essential to ensure the overall incompressibility condition of the stream function on $\Omega$ and thereby identify $u = \nabla \times \psi$. This requires satisfying appropriate boundary conditions, namely, the following combination of homogeneous Dirichlet/Neumann conditions for a computational cubic domain $\Omega = ]x_{\min}, x_{\max}[ \times ]y_{\min}, y_{\max}[ \times ]z_{\min}, z_{\max}[$:

$$\begin{cases} x = x_{\min} \text{ or } x = x_{\max} : \psi_y = \psi_z = \frac{\partial \psi_x}{\partial n} = 0, \\ y = y_{\min} \text{ or } y = y_{\max} : \psi_x = \psi_z = \frac{\partial \psi_y}{\partial n} = 0, \\ z = z_{\min} \text{ or } z = z_{\max} : \psi_x = \psi_y = \frac{\partial \psi_z}{\partial n} = 0. \end{cases}$$

Such boundary conditions ensure $(\nabla \times \psi) \cdot n = 0$, div $\psi = 0$ and $\psi \times n = 0$ at the same time, where $n$ is the normal field at the interface, and consequently lead to a zero average velocity field lifted by a prescribed flow rate $\overline{u}$ oriented in the flow direction. Indeed, given $u = \overline{u} + \nabla \times \psi$, one gets:

$$<u>_\Omega = \frac{1}{|\Omega|} \int_\Omega u \; dv = \overline{u} + \frac{1}{|\Omega|} \int_{\partial\Omega} \psi \times n \; ds = \overline{u},$$

which replaces the setting of the driving force $f$ by a prescribed flow rate, managed through the lifted vector $\overline{u}$.

Finally, using the stream function $\psi$ analytically ensures the divergence-free condition on the velocity as div $u =$ div $(\nabla \times \psi) = 0$. Overall, the velocity-vorticity formulation (13) of the DBS equation is subsequently coupled with the transport-reaction-diffusion PDE system developed in Section 2.2 to model reactive hydrodynamics at the pore-scale.

# 3 Method: hybrid grid-particle scheme (semi-Lagrangian)

The present work relies on a semi-Lagrangian numerical method, mixing Eulerian and Lagrangian formalism, to tackle dynamically evolving porous media due to reactive micrometric processes. Such a semi-Lagrangian approach has been successfully used for simulations of calcite dissolution at the pore scale in Etancelin et al. (2020) and extensively benchmarked against state-of-the-art numerical alternatives in Molins et al. (2021).

A Lagrangian formalism consists of describing the flow motion through the observation along time of a large number of fluid particles, with their attached intrinsic properties and spatially varying positions (Cottet and Koumoutsakos, 2000). Each particle is then tracked throughout the evolving mechanism (transport, diffusion, …) to measure variations in the main properties (velocity, concentration, …). On the contrary, from the Eulerian point of view, the previous property changes are characterized on a predetermined spatial grid along the dynamical process. This section is dedicated to presenting this hybrid formalism, which is subsequently improved to account for the heterogeneous diffusion of the chemical reactants through the porous matrix.

The particle formulation is especially well-suited for transport-dominant phenomena as it avoids the explicit discretization of convective terms and alleviates the consideration of their related stability constraints—namely, the CFL conditions which constrains the time step for a given spatial discretization. The lack of regularity in the particle distributions throughout the dynamic process is, however, a recurring problem of Lagrangian methods. Indeed, as the particle positions move according to the flow field gradients, accumulation or scarcity issues in the particle distribution commonly occur. This, thereby, requires periodic remeshing steps to avoid this problem and not to lose information: namely, one proceeds two successive interpolations from the disorganized particle structure to a regular grid and subsequently from the grid to the new particle distribution (Cottet and Poncet, 2004; Magni and Cottet, 2012).

This is particularly suitable for hybrid approaches, wherein dedicated solvers can be straightforwardly implemented in the Eulerian context before performing the remeshing step. This also allows a representation of the quantities of interest on the grid, which can be coupled with domain decomposition or mesh adaptation methods. Hybrid grid-particle formalism has, thereby, garnered considerable interest in addressing multiple complex phenomena in Computational fluid Dynamics (CFD) and geosciences (Beaugendre et al., 2012; Chatelain et al., 2008; Chatelin and Poncet, 2013; El Ossmani and Poncet, 2010; Hume and Poncet, 2021). Besides, incorporating high-order and compact support interpolation kernels makes it possible to reduce the overall computational complexity of the remeshing steps while keeping accurate predictions of the numerical solution. The choice of the interpolation kernels is, however, important to ensure a robust numerical method and guarantee properties such as mass conservation and sign-preservation of the interpolated quantities (Magni and Cottet, 2012). Improvements of the interpolation kernels, especially for applications to dissolution processes at the pore scale, have been investigated by Etancelin et al. (2020). Such improvements focused on sign preservation and accurate high-order interpolation through a correction step of the potential over-diffusive effects resulting from the remeshing step. This provides a well-established hybrid grid-particle framework that can robustly address pore-scale reactive flows.

In the present work, we aim to benefit from the main advantages of both approaches to model reactive hydrodynamics at the pore scale. We will, thereby, use a Lagrangian description for the chemical equations—including the heterogeneous diffusion operator and reactant transport—with an underlying regular grid for solving the DBS equation in its velocity-vorticity formulation. To do so, we describe in this sections this original numerical scheme, by detailing its components and how they are linked together:

- Section 3.1 describes the Lagrangian formalism of reaction-diffusion-transport equations and its resulting dynamical system (the particle formalism): it requires a velocity field to transport particles and the computation of a heterogeneous diffusion,
- Section 3.2 shows how this velocity, solution to the Darcy-Brinkman-Stokes, can be computed by an operator-splitting strategy using the algorithm 1/Brinkman penalization 2/Diffusion using improved PSE 3/Projection on divergence-free fields,
- Section 3.3 details the computational method for the required heterogeneous diffusion (that is to say space and time variable diffusion, including the space mapping based on the porosity field), and shows how its original formalism from Degond and Mas-Gallic (1989) is improved to become intrinsically second order.

## 3.1 Reactive dynamical system using Lagrangian formulation

In this section, we present the Lagrangian formulation dedicated to the resolution of the reactive dynamical system (Equation 8). detailed in Section 2.2, and more specifically to the transport-reaction-diffusion Equation 5.

We define a set of $N_p$ fluid particles covering the computational domain $\Omega$ and characterize as triplets $(C_i, x_i, v_i)_{i=1..N_p}$ of species concentrations $C_{i,k}$ ($k = 1..N_m$ indexing the $N_m$ chemical species), positions $x_i \in \Omega$ and volumes $v_i$, where $i$ refers to the particle index. This mathematically introduces the particle description, denoted $C^h$, of the concentration fields as follows:

$$C^h(t) = \sum_{i=1}^{N_p} C_i(t) \; v_i(t) \; \delta_{x_i}(t) \qquad (11)$$

which only depends on time $t$, and where $\delta$ refers to the Dirac function. The Lagrangian formulation of Equation 5 can then be written using the particle description Equation 11:

$$\begin{cases} \dfrac{dC_{i,k}}{dt} = R_k(\mathbf{C}_i(t)) + [\text{div}(\alpha_k(\varepsilon)\nabla^\varepsilon C_k)]_{x_i(t)} & \forall k = 1..N_m \\ \dfrac{dx_i}{dt} = [\varepsilon^{-1} u]_{x_i(t)} \\ \dfrac{dv_i}{dt} = 0 \end{cases}$$

given the incompressibility condition div $u = 0$ and the notations introduced in Section 2.2. This results in a dynamical system over the particles whose positions are controlled by the field $\varepsilon^{-1} u$, and volumes remain constant under divergence-free conditions. The main advantage of such a Lagrangian formulation (20) is that the transport term div($\varepsilon^{-1} u C_k$) vanishes along with its stability condition and, thereby, the method presents the ability to use arbitrary large time steps. This is, especially, of great interest when the Courant–Friedrichs–Lewy (CFL) condition on the transport term induces a stronger constraint on the time step compared to the diffusion stability condition.

The velocity field $u$ in (20) arises from the solution of the DBS equation which is solved on an underlying Cartesian grid and coupled with the Lagrangian formulation of the chemical PDE system. Regarding such a strong coupling between these equations, one needs to interpolate on the grid the particle description of the solid chemical species—namely, $C^h_{\text{CaCO}_{3(s)}}$ — which is related to the micro-porosity field $\varepsilon$ and consequently involved in the DBS model. Similarly, the velocity field subsequently needs to be interpolated on the particles to solve the Lagrangian set of chemical equations. This requires a convolution with high-order remeshing kernels with compact supports (Etancelin et al., 2020; Magni and Cottet, 2012). The dynamical system (20) is finally integrated using standard numerical methods for Ordinary Differential Equations (ODE), such as explicit Runge-Kutta, while the diffusion term div($\alpha_k(\varepsilon)\nabla^\varepsilon C_k$) is approximated through the Particle-Strength- Exchange (PSE) method, described in Section 3.3 and detailed in Supplementary Appendix.

In the present work, we incorporate in the semi-Lagrangian workflow the consideration of robustly estimating Archie's law term through such a particle-based PSE method. Thanks to this method, we make it possible to fully address the superficial gradient $\nabla^\varepsilon$ approximation with heterogeneous diffusion throughout the porous matrix.

## 3.2 Splitting operator strategy

The semi-Lagrangian formalism introduced in the previous Section 3.1 intrinsically relies on splitting strategies. Time-splitting methods, also known as fractional time-step algorithms, arise in many fields of computational science related to physics-based models and have been developed by Chorin (1973) in the context of vortex methods for the Navier-Stokes equation. Such methods have been widely investigated afterward to separate the resolution of distinct physical phenomena and render more efficient algorithms (Beale and Majda, 1981; Hume and Poncet, 2021; Faragó, 2008). Indeed, one of the main advantages of splitting strategies is one can use different approaches for the distinct parts of the overall model, namely, either a Lagrangian or Eulerian formulation. This also straightforwardly extends to the choice of the numerical solver available for each component, allowing the use of the most efficient, accurate, and robust schemes independently.

The first natural splitting arising in workflow, thereby, lies in the semi-Lagrangian formalism itself, wherein we do not consider solving the overall Partial Differential Equation (PDE) system at once. We rather separate the transport-diffusion-reaction dynamics in its Lagrangian formulation from the pore-scale hydrodynamic resolved on the underlying Cartesian grid. The hydrodynamic part, composed of the DBS equation in the velocity-vorticity formulation, is also solved through a time-splitting method. Indeed, we approximate the solution of Equation 9 by the limit in time of the evolution equation

$$\frac{\partial \omega}{\partial t} - \mu \Delta \omega + \mu \kappa_b^{-1} \nabla \times \left( \frac{(1-\varepsilon)^2}{\varepsilon^2} u \right) = 0,$$

together with $\omega = \nabla \times u$, using a three-step operator splitting strategy coupled with a fixed-point algorithm. This means that considering a sequence $(u_m, \omega_m)$ of velocity-vorticity we aim to successively perform, over a time interval $[t_m, t_{m+1}]$ with $t_m = m\delta t$, Brinkman penalization, diffusion, and projection on divergence-free fields. The latter is achieved through the reconstruction of the velocity field $u$ based on the stream function $\psi$ (see Section 2.3). In practice, these three steps are specifically defined as:

- The Brinkman iteration given by the ordinary differential equation $\frac{\partial u}{\partial t} + \mu\lambda(\varepsilon)(u+\overline{u}) = 0$ with prescribed flow rate $\overline{u}$ and $\lambda(\varepsilon) := \kappa_b^{-1}(1-\varepsilon)^2\varepsilon^{-2}$, whose exact solution after a $\delta t$ is generated by

$$\Lambda(u) := e^{-\mu\lambda(\varepsilon)\delta t}(u+\overline{u}) - \overline{u}$$

- The vorticity diffusion iteration, $\frac{\partial \omega}{\partial t} - \mu\Delta\omega = 0$, solved using an implicit Euler scheme given by the operator

$$\mathcal{D}_\omega(u) := [I - \mu\delta t\Delta]^{-1}(\nabla \times u)$$

- The projection step $\Pi(\zeta) = \nabla \times ((-\Delta^{-1})\zeta)$ which takes as $\zeta$ the right-hand side of the Poisson equation $-\Delta\psi = \zeta$ satisfying the boundary conditions (17), and followed by $u = \nabla \times \psi$.

Overall, this leads to the full iteration of the Brinkman-Diffusion-Projection splitting $\Pi \circ \mathcal{D}_\omega \circ \Lambda$ over a time step $[t_m, t_{m+1}]$, which reads as follows:

$$u_{m+1} = \Pi \circ \mathcal{D}_\omega \circ \Lambda(u_m)$$

and whose consistency has been theoretically discussed in Hume and Poncet (2021). One should notice that this projection step is not a projection by pressure gradient correction as in Chatelin and Poncet (2013), but an operator that takes the vorticity field $\omega$ and retrieves a divergence-free velocity field whose mean velocity is zero. The final velocity field is subsequently given by

$$u = \overline{u} + \lim_{m \to \infty} u_m$$

whose average is $\overline{u}$ and satisfies the Kozeny-Carman relation inside the solid region.

From a numerical perspective, we consider the exact treatment of the Brinkman term, a fourth-order finite difference scheme for the discrete curl operator, and FFT solvers for the vorticity diffusion and stream function recovery. Using FFT avoids matrix assembly procedure and, therefore, consists of efficient solvers in terms of computational time and memory storage requirements. Besides, the complexity of such algorithm scales as $\mathcal{O}(N_p \log(N_p))$, where $N_p$ is recalled to refer to the number of particles. Finally, a stopping criterion on this fixed point algorithm is also defined based on the relative residual norm on the velocity, which manages the convergence of the pore-scale hydrodynamics toward a stationary state.

The updated velocity $u_{m+1}$ is subsequently interpolated from the grid to the particles and used for solving the Lagrangian reactive system, which is split into convective/remeshing and diffusive/reactive steps. Regarding the convection, the particle trajectories are pushed to the next step through a directional advection, given the field $\varepsilon^{-1}u$ according to the Lagrangian formulation (20), and are then remeshed to avoid stagnation issues. The purpose of such a directional splitting is to reduce the dimensionality of the overall advection problem by considering several one-dimensional equations, and is developed in Cottet et al. (2014); Magni and Cottet (2012). The particle remeshing is also addressed using directional treatment of the multidimensional convolution. This means that within a time step $[t_m, t_{m+1}]$, the joint step of advection/remeshing of the particles is successively performed by alternating the spatial directions (Etancelin, 2014; Keck, 2019). This presents the advantage of significantly improving the computational efficiency of the method and allows the use of high-order remeshing kernels with large stencils while maintaining a minimal cost compared to multidimensional cases. This is also well-suited to parallel implementation on GPU architecture. The dimensional splitting is addressed, in practice, by a second-order Strang formula (Strang, 1968) and coupled with a second-order Runge-Kutta for time integration. The diffusion/reaction step is finally solved by means of a second-order explicit Runge-Kutta scheme along with PSE approximation of the heterogeneous diffusion operator. Once the Lagrangian formulation of the chemistry has been fully updated and remeshed on the underlying grid, one starts pushing the DBS hydrodynamics to the next sequential step of these temporal iterations.

Such an operator splitting strategy, in the context of a semi-Lagrangian approach, has been applied to the modeling of dissolution processes on a 2D synthetic calcite core and validated against state-of-the-art alternatives and experiments in Molins et al. (2021). This has also been used in Etancelin et al. (2020) on real porous media structures at the pore scale to investigate the dissolution of 3D carbonate rocks arising from $\mu$CT scans. Nonetheless, these previous works assumed that the superficial gradient $\nabla^\varepsilon$ involved in the heterogeneous diffusion could be approximated by the gradient operator, and subsequently addressed this Archie's law term with standard finite differences schemes. In Section 3.3, we intend to alleviate this assumption and, therefore, improve the semi-Lagrangian method by incorporating in the workflow a PSE approximation of the heterogeneous diffusion.

## 3.3 Particle-strength-exchange method and Archie's law approximation

The PSE method consists in the approximation of a diffusion operator $\text{div}(\mathbf{L}\nabla f)(x)$ with $x \in \Omega \subset \mathbb{R}^d$ and $\mathbf{L}$ a positive symmetric matrix, accounting for heterogeneous diffusion for instance. The main idea is then to approximate the diffusion by an integral operator, more suitable for particle methods:

$$Q^\xi \cdot f(x) = \int_\Omega \sigma^\xi(x,y)(f(y) - f(x))\,dy$$

where the kernel $\sigma^\xi$ is supposed to be symmetric and satisfies some moment conditions, described thereafter and detailed in Supplementary Appendix. In the Lagrangian formulation, a particle approximation of the function $f$, denoted $f^h$, is also introduced based on the particle triplet $(f_i, x_i, v_i)$, such that:

$$f^h = \sum_{i=1}^{N_p} f_i\, v_i\, \delta_{x_i} \quad \text{where} \quad f_i = f(x_i)$$

where $x_i$ and $v_i$ are respectively the particle positions and volumes, while $\delta_{x_i}$ refers to the Dirac measure at position $x_i$. With such a $N_p$-particle representation of the function $f$, a discrete version of the operator $Q^\xi$ is obtained by using the particles as quadrature points where $h$ refers to the characteristic distance between particles. This results in the following quadrature expression, called Particle-Strength-Exchanges:

$$Q^\xi \cdot f^h(x_k) = \sum_{x_l \in \mathcal{S}(x_k)} \sigma^\xi(x_k, x_l)(f_l - f_k)\, v_l.$$

where $\mathcal{S}(x_k) \coloneqq \text{Supp}(\sigma^\xi(x_k, \cdot))$ refers to the set of points in the support of the kernel function $\sigma^\xi$.

In Supplementary Appendix, we detail that we can explicitly compute two constants $\gamma_1$ and $\gamma_2$ based on a given scale $\xi$ and a given function $\Theta$ (named stencil generator) such as

$$\sigma^\xi(x,y) = \frac{1}{\xi^{d+4}} \Theta\left(\frac{y-x}{\xi}\right) \frac{\mathbf{m}(x) + \mathbf{m}(y)}{2} : (x-y)^{\otimes 2}$$

with

$$\mathbf{m} = c_0\, \mathbf{L} - c_1\, \text{Tr}(\mathbf{L})\,\text{Id}_3 + \mathbf{H}$$

and

$$c_0 = \frac{2(\gamma_1 + 2\gamma_2)}{\gamma_1^2 + \gamma_1\gamma_2 - 2\gamma_2^2},\ c_1 = \frac{2\gamma_2}{\gamma_1^2 + \gamma_1\gamma_2 - 2\gamma_2^2},\ \text{and}$$

$$H_{ij} = \left(\frac{\gamma_1^2 - \gamma_1\gamma_2 - 6\gamma_2^2}{\gamma_2(\gamma_1^2 + \gamma_1\gamma_2 + 2\gamma_2^2)}\right)(1 - \delta_{ij})L_{ij}$$

where $\delta_{ij}$ is the Kronecker symbol, $\gamma_1$, and $\gamma_2$ are the coefficients based on the fourth moments of $\Theta$ defined in 3D by

$$\gamma_1 = \sum_{x \in \mathbb{J}} x_k^4 \Theta(x) h^3, \quad k \in [1,3]$$
$$\gamma_2 = \sum_{x \in \mathbb{J}} x_k^2 x_l^2 \Theta(x) h^3, \quad k \neq l \in [1,3]$$

for $\mathbb{J} \subset h\mathbb{Z}^3$ a three-dimensional lattice, including at least one neighborhood of the current mesh point. The convergence of the original PSE method introduced in Degond and Mas-Gallic (1989) depends on the scale $\xi$ as $\mathcal{O}[(h/\xi)^2]$, which becomes $\mathcal{O}(1)$ when $\xi$ is adapted linearly to $h$. The present discrete corrected version developed in Poncet (2006) and Schrader et al. (2010) is intrinsically second order convergence $\mathcal{O}(h^2)$ and is detailed in Supplementary Appendix.

This computational method of diffusion is used to compute the Archie's law, involving a tortuosity index $\eta$ in the heterogeneous diffusion operator

$$\mathcal{D}(\varepsilon, C) \coloneqq \text{div}(\varepsilon^{1+\eta}\nabla(\varepsilon^{-1}C))$$

Involved in the PDE system (11). In this case, one gets a particular expression for the diffusion matrix $\mathbf{L} = \varepsilon^{1+\eta}\, \mathbf{I}_3$ with $\mathbf{I}_3$ the identity matrix in $\mathbb{R}^3$. This finally leads to the following discretized Archie's law diffusion operator (30) using the renormalized PSE scheme:

$$Q^\xi \cdot f^h(x_k) = \frac{1}{\xi^7} \sum_{l \sim k} (f_l - f_k) \Theta\left(\frac{x_l - x_k}{\xi}\right) \frac{(c_0 - 3c_1)}{2}$$
$$\times (\varepsilon^{1+\eta}(x_k) + \varepsilon^{1+\eta}(x_l)) |x_l - x_k|^2 v_l \quad (12)$$

where $|.|$ is the Euclidean norm in $\mathbb{R}^3$ and $f_\bullet \coloneqq \varepsilon^{-1}(x_\bullet)C(x_\bullet)$. The overall Formula 12 accounts for the heterogeneous diffusion and ensures the accurate management of the chemical reactant penetration, given by its concentration field $C$, within the porous matrix. Indeed, one of the main advantages of the PSE scheme is that it includes all the lattice neighborhoods in the computation of the heterogenous diffusion, unlike crossed finite differences scheme. Finally, this guarantees the strict conservation of the reactant exchanges between the fluid portion and the porous matrix, and is numerically validated at the second order in Section 4 of Supplementary Appendix.

## 4 Application to precipitation and crystallization

$CO_2$ mineral storage in natural underground reservoirs, such as saline aquifers, involves competing geochemical phenomena

occurring at a large variety of scales. Among them, mineral dissolution and precipitation play crucial roles. On one side, studying the dissolution of native carbonate species, already present in the aquifers, provides insight into potential leakage issues and queries the reservoir safety. On the other side, $CO_2$ trapping under the form of carbonate precipitates and crystals informs on the storage capacities of the reservoir. These geochemical processes also induce changes in the macro-scale properties of the subsurface material, including permeability and porosity evolutions, that need to be investigated to ensure sustainable management of the reservoir structures.

In this section, we develop mathematical models for calcite precipitation and crystallization at the pore scale, with special considerations on the reaction rate expressions arising in the PDE system (11). Numerical simulations are performed within the HySoP platform (Etancelin et al., 2022) along with PSE treatment of the heterogeneous diffusion on accelerated GPU devices and address porous sample arising from X-ray $\mu$CT observations. This enables the investigations of macro-scale property changes along the $CO_2$ mineral trapping on real 3D rock geometries, which is an important component in the overall study of $CO_2$ storage.

## 4.1 The transition state theory: from dissolution to precipitation modeling

Dissolution of the injected $CO_2$ in the aqueous phase of deep underground reservoirs will affect the pH of the formation water through the following series of chemical reactions:

$$CO_{2\,(g)} \rightleftharpoons CO_{2\,(aq)} \tag{13a}$$

$$H_2O + CO_2\,(aq) \rightleftharpoons H_2CO_3 \tag{13b}$$

$$H_2CO_3 \rightleftharpoons H^+ + HCO_3^- \tag{13c}$$

$$HCO_3^- - \rightleftharpoons H^+ + CO_3^{2-}. \tag{13d}$$

Indeed, once the $CO_2$ has dissolved into water and established a first equilibrium under the form of the weak acid $H_2CO_3$, the $H_2CO_3$ species dissociates successively to bicarbonate $HCO_3^-$ and carbonate $CO_3^{2-}$ ions as the pH increases. These chemical reactions are pH-dependent, and the distribution evolutions of all these carbonate species are displayed in Figure 2 against the pH of the solution. In alkaline media, the chemical reactions (Equation 13c) and (Equation 13d) can, therefore, join together to read as follows:

$$H_2CO_3 + 2OH^- \rightleftharpoons CO_3^{2-} + 2H_2O$$

such that carbonate ions are the main carbonate species present in the solution. Such transformations in the ionic species composition of the aquifer water will considerably impact the original mineral structure through chemical rock-water interactions such as carbonate dissolution and precipitation.

Historically, the dissolution and precipitation kinetics of calcite in the context of $CO_2$ injection have been studied since the 1970s, both from the experimental and theoretical sides.

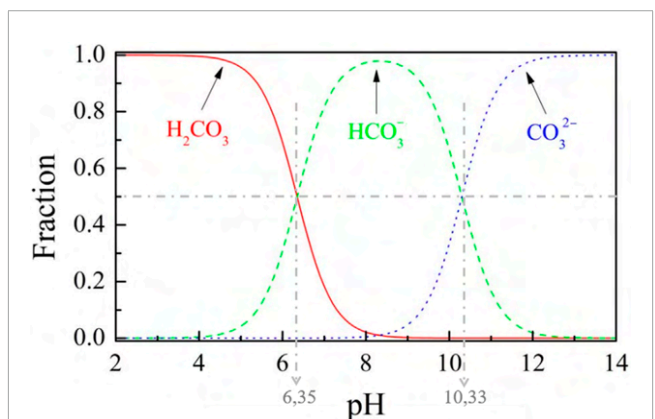

FIGURE 2
Distribution diagram of aqueous carbonate species against pH solution from Bohn et al. (1980) and Turk et al. (2015). Species distributions are represented as a fraction of total dissolved carbonate. The grey dotted lines highlight the transition pH of the chemical equilibria from (Equation 13).

Plummer et al. (1978) investigated the influence of several parameters on the forward reaction rates of calcite dissolution under far-from-equilibrium conditions. Among these parameters, one finds the partial pressure of $CO_2$ denoted $P_{CO_2}$, the hydrogen ions activities denoted $a_{H^+}$ — directly related to the pH — and the temperature. Their experimental work was subsequently extended by Chou et al. (1989) using a fluidized bed reactor to compare the dissolution kinetics mechanisms between various carbonate minerals—involving *inter alia* calcite, aragonite, and dolomite—at 25°C. One should notice that these experiments were conducted under laboratory conditions in terms of pressures and temperatures, in opposition to the current abilities of *in situ* experiments to manage realistic reservoir conditions (Andrew et al., 2013; Wigand et al., 2008). Nonetheless, these experimental studies have highlighted three kinetic mechanisms occurring simultaneously in the process of calcite dissolution due to $CO_2$ injection. Such mechanisms are given by the following chemical reactions:

$$CaCO_{3(s)} + H^+ \underset{}{\overset{K_1}{\rightleftharpoons}} Ca^{2+} + HCO_3^- \tag{14a}$$

$$CaCO_{3(s)} + H_2CO_3 \underset{}{\overset{K_2}{\rightleftharpoons}} Ca^{2+} + 2HCO_3^- \tag{14b}$$

$$CaCO_{3(s)} \underset{K_{-3}}{\overset{K_3}{\rightleftharpoons}} Ca^{2+} + CO_3^{2-} \tag{14c}$$

where the notations $K_i$, $i = 1\ldots 3$ refer to forward reaction rate constants, depending on the temperature (Busenberg and Plummer, 1986; Plummer et al., 1978; Plummer and Busenberg, 1982), and $K_{-3}$ is the backward reaction rate corresponding to the reverse calcite precipitation process in Equation 14c. They experimentally identified both the forward and backward reaction rates and established the validity of kinetic models for carbonate dissolution and precipitation in comparison to thermodynamics theoretical considerations. In the present study, we assume the production of pure calcite without any of its polymorphs like aragonite or vaterite.

Meanwhile, mineral reaction rates were, indeed, theoretically investigated by Lasaga (1981) using the Transition State Theory

(TST), originally formulated by Eyring (1935). Since then, this formalism has successfully been extended (Aagaard and Helgeson, 1982; Lasaga, 1984; Steefel and Lasaga, 1994) and widely accepted in current kinetic geochemical models (Etancelin et al., 2020; Molins et al., 2012; Steefel et al., 2015). In this context, the reaction rates are commonly expressed as the product of far-from-equilibrium terms, involving the activities of the chemical species in solution, with an affinity term written as a function of the Gibbs free energy change $\Delta G$ for close to the equilibrium conditions. Considering the chemical model of calcite dissolution (34) suggested by Plummer et al. (1978) and Chou et al. (1989) and the vector of concentrations **C**, the reaction rate arising from TST writes:

$$R(\mathbf{C}) = A_s \left( K_1 a_{H^+} + K_2 a_{H_2CO_3} + K_3 \right) \left( \frac{a_{Ca^{2+}} \, a_{CO_3^{2-}}}{K_{eq}} - 1 \right) \quad (15)$$

where $K_{eq}$ is the equilibrium constant of the reaction, also called the solubility product, $A_s$ is the reactive surface area of the mineral—in $m^{-1}$. The notations $a_\bullet = \gamma_\bullet C_\bullet$ refer to the dimensionless species activities with $\gamma_\bullet$ and $C_\bullet$, respectively, their activity coefficients and molar concentrations—whose unit is mol.m$^{-3}$. It follows that the micro-porosity changes reads as:

$$\frac{\partial \varepsilon}{\partial t} = -\upsilon R(\mathbf{C}),$$

given Equation 6 from Section 2.2 and the relation $C_{CaCO_{3(s)}} = (1-\varepsilon)/\upsilon$. Denoting by $Q = a_{Ca^{2+}} \, a_{CO_3^{2-}}$ the ion activity product, one obtains the following relation between $Q$ and the Gibbs energy change (Aagaard and Helgeson, 1982; Lasaga, 1984):

$$\Delta G = RT \ln \left( \frac{Q}{K_{eq}} \right)$$

with $T$ the temperature in Kelvin K, and $R$ the universal gas constant in J.mol$^{-1}$.K$^{-1}$. The sign of the reaction term $R(\mathbf{C})$ in (Equation 15) is driven by the sign of $\ln(Q/K_{eq})$ that is negative for dissolution and positive for precipitation, which is consistent with the convention from Section 2.2.

From now on, we focus on the concern of calcite precipitation and crystallization resulting from $CO_2$ injection based on the series of homogeneous reactions (Equation 13) along with the mineral-solute interaction given by Equation 14c. In practice, we enforce a pH greater than 10.33 such that the carbonate ions $CO_3^{2-}$ are the predominant species (see Figure 2). This enables the restriction of the overall set of chemical reactions (Equation 13) to merely consider the Equation 14c in the sense that we assume the intermediate reactions as instantaneous and conservative—without loss of quantity of matter. Such an assumption is acceptable, in practice, since fluid-mineral reaction rates are usually slower than intra-aqueous reaction rates. Therefore, the initial concentration of carbonate ions, denoted $C_{CO_3^{2-}}(x, t=0)$ for $(x,t) \in \Omega \times [0, T_f]$ following the notations introduced in Section 2, is directly related to the partial pressure of injected $CO_2$ by means of the Henry law. The latter states, at a constant temperature, the relation between the amount of dissolved gas in a solute and its partial pressure based on Henry's law constant denoted $K_H$, which depends on the gas and temperature, such that for the $CO_2$ at 25°C one gets:

$$C_{CO2(aq)} = \frac{P_{CO_2}}{K_H} \simeq C_{CO_3^{2-}}(x, t=0) \quad (16)$$

where $K_H = 29.41$ L.atm.mol$^{-1}$. Considering such an alkaline medium—with pH > 10.33 — also results in the treatment of the chemical reaction (Equation 14c) as completely irreversible which corresponds to far-from-equilibrium conditions modelling the calcite precipitation chemical reaction:

$$Ca^{2+} + CO_3^{2-} \xrightarrow{K_{-3}} CaCO_{3(p)} \quad (17)$$

where the subscript $p$ here refers to the precipitate form of the calcite product. In this case, the rate constants $K_1 = K_2 = 0$ in (35) and the affinity term dependent on the Gibbs energy satisfies the condition $\ln(Q/K_{eq}) \gg 0$ — corresponding to a supersaturated solution—so that we obtain an overall reaction rate for the calcite precipitation which reads as:

$$R_{prec}(\mathbf{C}) = K_{-3} \, A_s \, a_{Ca^{2+}} \, a_{CO_3^{2-}} \quad (18)$$

where $K_{-3} = K_3/K_{eq}$, which theoretically results from the TST law in Equation 15 and has been experimentally validated, *inter alia*, by Chou et al. (1989).

Therefore, in the following, we rely on the kinetic formulation of the mineral precipitation given by Equation 18, considering the rate laws determined by laboratory experiments (Chou et al., 1989) and normalized by the reactive surface area of the mineral $A_s$ (Plummer et al., 1978). As the geometry evolves, the micro-porosity $\varepsilon$ and the reactive specific area $A_s$, associated with the porous structure, also change. These evolutions are taken into account in the reaction rates management and the hydrodynamic modeling of the reactive process (see the overall PDE system (Equation 8) in Section 2.2). The calcite precipitation reaction is subsequently supplemented with a crystallization model which is elaborated in the next Section 4.2.

## 4.2 Crystal growth modeling: a two-step process

Crystal growth kinetics involves complex mechanisms occurring simultaneously and depending, *inter alia*, on the concentration of the constituent ions in the solute, but also on attachment frequencies of the ions or precipitates to lattice growth sites (Nielsen and Toft, 1984; Wolthers et al., 2012). Indeed, the growth rate is first controlled by advection and diffusion of the $Ca^{2+}$ and $CO_3^{2-}$ ions to the crystal surface coupled with a surface adsorption process that largely hinges on the crystal lattice shape. For instance, the growth of crystal aggregates is more likely to occur near kinks or corners (Nielsen and Toft, 1984; Yoreo and Vekilov, 2003). Mineral heterogeneity of the pore interface is also an important factor that influences the crystal growth location and morphology, providing preferential sites (Lioliou et al., 2007). This first process is commonly called primary heterogeneous nucleation, for which the crystallization reaction is catalyzed by the solid surface of the porous medium. In the absence of solid interface, crystal clusters can also form spontaneously in the solute, which is known as primary homogeneous nucleation and is closely related to the supersaturation state of the solution in order to initiate the nucleation—namely, satisfying the condition $\ln(Q/K_{eq}) \gg 0$. Finally, secondary nucleation occurs in the presence of existing crystals and is more likely to generate large crystal aggregates at the mineral surface. Overall, calcite crystallization results from a combination of all these previous phenomena.

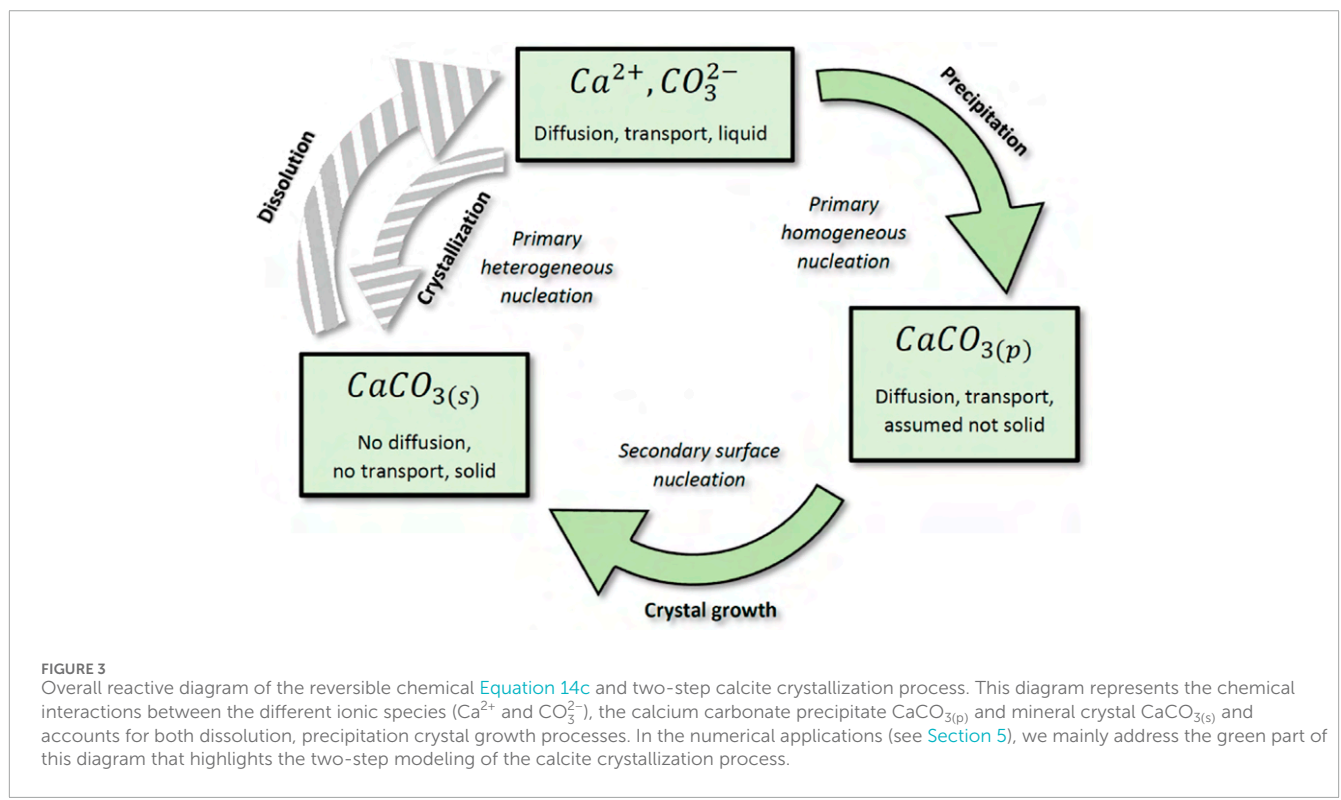

FIGURE 3
Overall reactive diagram of the reversible chemical Equation 14c and two-step calcite crystallization process. This diagram represents the chemical interactions between the different ionic species ($Ca^{2+}$ and $CO_3^{2-}$), the calcium carbonate precipitate $CaCO_{3(p)}$ and mineral crystal $CaCO_{3(s)}$ and accounts for both dissolution, precipitation crystal growth processes. In the numerical applications (see Section 5), we mainly address the green part of this diagram that highlights the two-step modeling of the calcite crystallization process.

In this section, we consider a two-step crystallization process wherein calcite precipitates, also referred to as nuclei and denoted $CaCO_{3(p)}$, are first generated within the solute during the so-called nucleation stage according to the chemical Equation 17. These precipitates are subsequently aggregated at the mineral surface through adsorption phenomena during the crystal growth step. This sequential crystallization process is described in Figure 3, where the notation $CaCO_{3(s)}$ stands for the calcite crystal. In the applications developed in Section 5, we consider that the solid matrix of the 3D porous sample has a similar carbonate nature to the calcite crystal generated, though rock mineral heterogeneities can be integrated into the numerical framework as prospects. From now on, we refer to precipitation as the primary homogeneous nucleation and we investigate the surface attachment of the calcite precipitate, $CaCO_{3(p)}$, based on an autocatalytic process to model calcite crystal growth, referred to as the secondary surface nucleation in Figure 3. In the reaction scheme from Figure 3, we account for the precipitate diffusion and advection until the solid boundary where it leads to crystal growth through adsorption phenomena. Besides, we neglect the direct crystallization process induced by the solute diffusion to the solid matrix and the so-called primary heterogeneous nucleation.

In the literature, two distinct approaches are mainly developed when considering precipitation and crystal growth modeling altogether. The former can be regarded as "deterministic models" relying on the TST modeling developed in Section 4.1. Noiriel et al., for instance, investigated the effects of pore-scale precipitation on permeability through a combination of X-ray μCT experiments and "deterministic" modeling (Noiriel et al., 2021). They also derived crystal growth rates directly from the μCT through an imaging comparison between the beginning and end of the precipitation experiment. In this case, only two μCT scans were performed, and therefore, the process should be understood as distinct from 4D μCT experiments incorporating time dynamics. Such experimental identification of crystal growth rates can, however, be prone to intrinsic imaging limitations (Perez et al., 2022) and result in wide discrepancies in the reaction rate estimations. Nonetheless, their results showed satisfactory agreement between the experiments and numerical experiments for precipitation processes into fractures (Noiriel et al., 2021). Alternative modeling approaches lie in the probabilistic nature of nucleation or crystal growth and are referred to as "probabilistic models" (Fazeli et al., 2020; Masoudi et al., 2021; Nooraiepour et al., 2021a). Wolthers et al., for instance, developed a probabilistic approach for calcite crystal growth based on the nature of the kink sites depending on their ionic affinities and attachment frequencies of the constituent ions (Wolthers et al., 2012). Estimations of such adsorption frequency ranges can also be found in the literature (Christoffersen and Christoffersen, 1990; Nielsen, 1984). Finally, while it is commonly established experimentally that crystal growth occurs preferentially at kinks and corners of the surface lattice, few models incorporate the geometrical dependency of the crystal aggregation in the reaction rates (von Wolff et al., 2021).

In the present article, we propose a new approach coupling a deterministic model for the precipitation, which directly depends on the supersaturation ratio following the TST formalism, and a probabilistic formulation of the crystal growth process. The latter accounts for the adsorption frequencies of the precipitate to the growth sites with a coefficient, quantifying the physical probability of attachment rate, denoted $P_{ad}$, which relies on a locally averaged mineral volume fraction. In this formalism, one obtains the crystal growth reaction rate, which is expressed as:

$$R_{crys}(\mathbf{C}) = K_c \ P_{ad} \ C_{CaCO_{3(p)}} \quad (19)$$

with $K_c$ the adsorption frequency (expressed in s$^{-1}$). It possible then to include the geometrical dependency in the crystal growth reaction rate by means of involving the neighborhood of the solid using a convolution of the porosity (the size of the neighborhood is provided by the support of the convolution kernel). Practically, this leads to the relation

$$P_{ad} = C(1 - \varepsilon * W_h)\varepsilon^m, \quad (20)$$

where $W_h(x) = h^{-d}W^{\otimes d}(x/h)$ is the rescaled kernel based on a local averaging kernel $W$, so that $(1 - \varepsilon * W_h)\varepsilon$ is enhanced in the layer close to the fluid/solid interface and depending on the solid proportion in the neighborhood (see Figures 4, 5). The calibration of $C$ and kernel support size $h$ can be established by comparison with experiments (Poonoosamy et al., 2023) or numerical simulations in standardized geometry (Varzina et al., 2020; Patel et al., 2021), or estimated in the same spirit as the *crystallites* used in the LBM schemes in Masoudi et al. (2024).

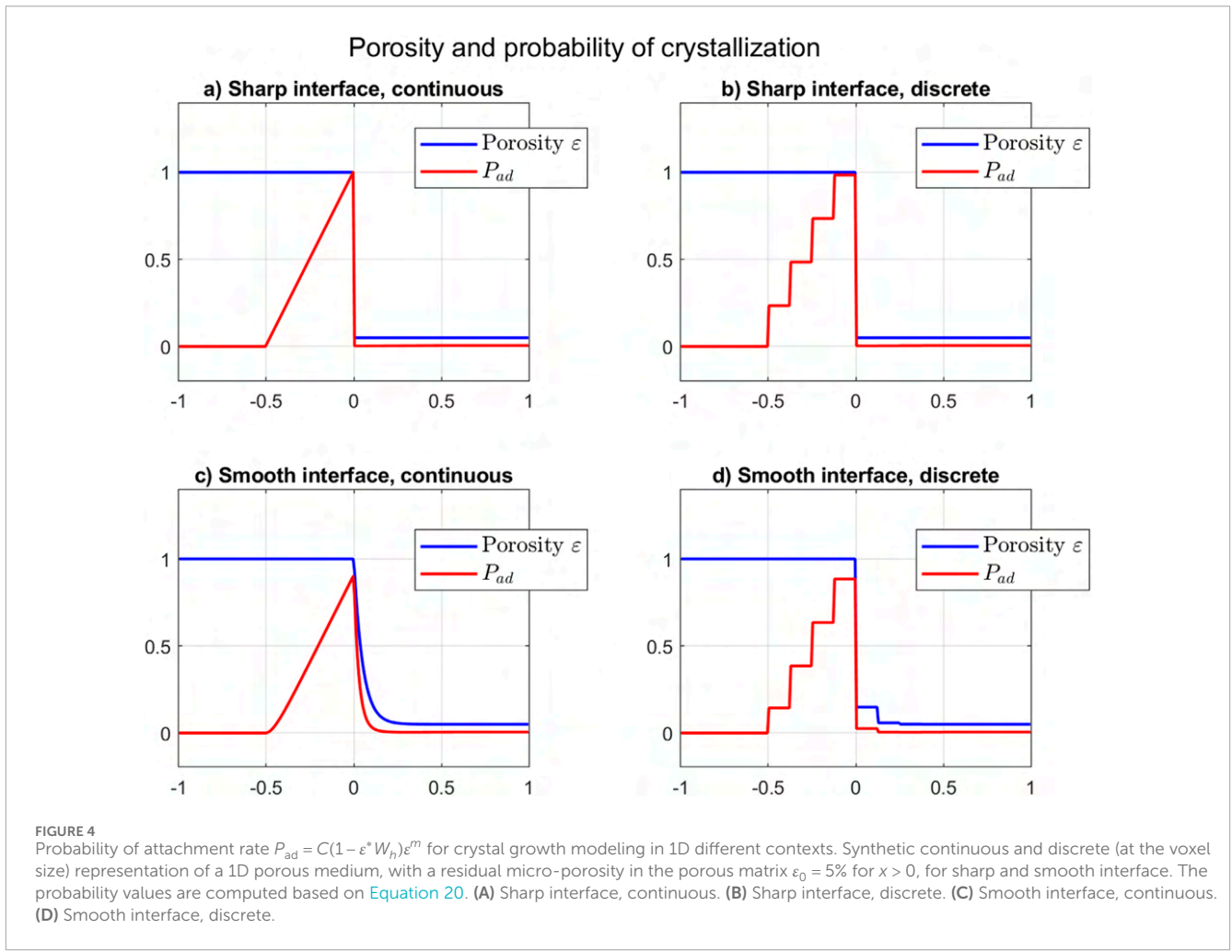

FIGURE 4
Probability of attachment rate $P_{ad} = C(1 - \varepsilon * W_h)\varepsilon^m$ for crystal growth modeling in 1D different contexts. Synthetic continuous and discrete (at the voxel size) representation of a 1D porous medium, with a residual micro-porosity in the porous matrix $\varepsilon_0 = 5\%$ for $x > 0$, for sharp and smooth interface. The probability values are computed based on Equation 20. **(A)** Sharp interface, continuous. **(B)** Sharp interface, discrete. **(C)** Smooth interface, continuous. **(D)** Smooth interface, discrete.

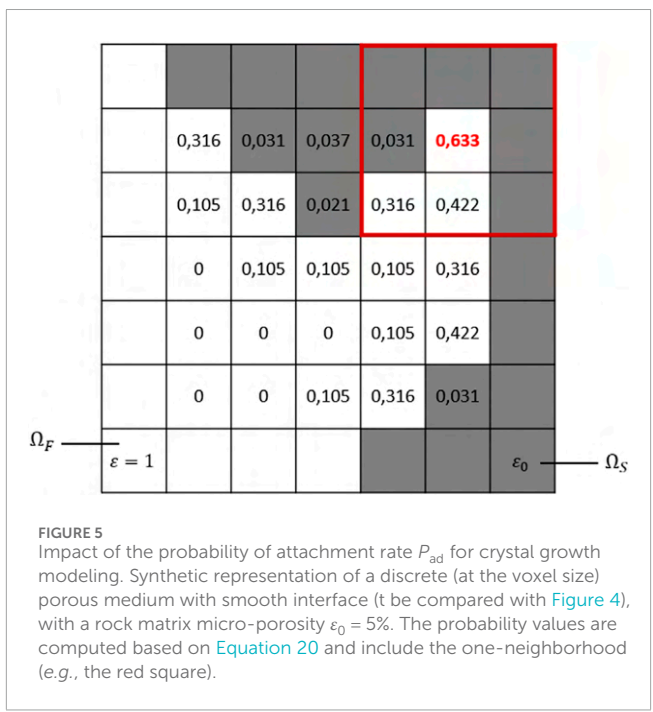

FIGURE 5
Impact of the probability of attachment rate $P_{ad}$ for crystal growth modeling. Synthetic representation of a discrete (at the voxel size) porous medium with smooth interface (t be compared with Figure 4), with a rock matrix micro-porosity $\varepsilon_0 = 5\%$. The probability values are computed based on Equation 20 and include the one-neighborhood (e.g., the red square).

In the case of $K_c$ being a production of crystal volume per unit of volume and time, then the normalization coefficient $C$ satisfies $1/C = \langle (1 - \varepsilon^* W_h)\varepsilon^m \rangle_\Omega$ so that $P_{ad}$ is a probability distribution, and the index $m$ describes the ability to decrease strongly the reaction rate inside the solid in the spirit of the Archie's law (in practice $m = 2$ is suitable). In the case of $K_c$ being the local production of crystal at the fluid/solid interface, which is the case considered in the present study, then $C$ satisfies

$$1/C = (1 - \varepsilon_0)/2$$

so that $P_{ad}$ is a point-wise probability of capture by the crystal. This is based on the jump between $\varepsilon_0$ and 1 through a sharp interface, with a factor 1/2 (half the integral of the kernel $W$, which can be slightly adjusted for concave interface as shown in Poncet (2007)). This convolution-based formulation is appropriate for a crystallization process and is inspired by the gradient-based technique from Luo et al. (2012) and Soulaine et al. (2017) that locates the first layer on the solid side and is suitable for dissolution processes. In practice, the local averaging kernel $W$ can be as simple as $W = 1_{[-1,1]}/2$ or Gaussian in a continuous description, or approximated by its discrete formulation $W = (\delta_{-1} + \delta_0 + \delta_1)/3$, or even with weighting in order to modulate the length of capture with respect to the grid size.

The Figure 4 shows different probability distributions in 1D with a residual micro-porosity of $\varepsilon_0 = 5\%$ (in the area $x > 0$), for sharp and smooth fluid/solid interface, and for the continuous and discrete formulations described above. The resulting probability values in 2D are represented in Figure 5 for a synthetic example with smooth interface and discrete formulation (to be compared to the 1D version of Figure 4D), where the residual micro-porosity in $\Omega_S$ is estimated to $\varepsilon_0 = 5\%$ as well.

## 4.3 Final system of PDEs

Finally, we define the vector of concentrations

$$\mathbf{C} = \left(C_{CaCO_{3(s)}}, C_{CaCO_{3(p)}}, C_{CO_3^{2-}}, C_{Ca^{2+}}\right)$$

and consider the reactions rates $R_{prec}(\mathbf{C})$ and $R_{crys}(\mathbf{C})$ respectively given by Formula 40, 41. Overall, the calcite crystallization modeled as a two-step process of precipitation and crystal growth, according to the reaction scheme from Figure 3, writes:

$$\begin{cases} -\text{div}(2\mu D(u)) + \mu \kappa_b^{-1} \frac{(1-\varepsilon)^2}{\varepsilon^2} u = \varepsilon(f - \nabla p), & \text{in } \Omega \times ]0, T_f[ \\ \frac{\partial C_{CO_3^{2-}}}{\partial t} + \text{div}\left(F\left(C_{CO_3^{2-}}\right)\right) = -R_{prec}(\mathbf{C}), & \text{in } \Omega \times ]0, T_f[ \\ \frac{\partial C_{Ca^{2+}}}{\partial t} + \text{div}\left(F\left(C_{Ca^{2+}}\right)\right) = -R_{prec}(\mathbf{C}), & \text{in } \Omega \times ]0, T_f[ \\ \frac{\partial C_{CaCO_{3(p)}}}{\partial t} + \text{div}\left(F\left(C_{CaCO_{3(p)}}\right)\right) = R_{prec}(\mathbf{C}) - R_{crys}(\mathbf{C}), & \text{in } \Omega \times ]0, T_f[ \\ \frac{\partial C_{CaCO_{3(s)}}}{\partial t} = R_{crys}(\mathbf{C}), & \text{in } \Omega \times ]0, T_f[ \\ \varepsilon = 1 - \nu C_{CaCO_{3(s)}}, & \text{in } \Omega \times ]0, T_f[ \\ F(C_\bullet) = \varepsilon^{-1} u C_\bullet - \alpha_\bullet(\varepsilon) \nabla^\varepsilon C_\bullet, \quad \alpha_\bullet(\varepsilon) = \varepsilon^\eta D_{m,\bullet}, & \text{in } \Omega \times ]0, T_f[ \\ + \text{ adequate boundary and initial conditions, along with div } u = 0 \end{cases}$$
(21)

where the advective and diffusive flux $F(C_\bullet)$ is taken from expressions in Section 2.2, using the lifted gradient $\nabla^\varepsilon(\cdot) = \varepsilon \nabla(\varepsilon^{-1} \cdot)$.

The reactive hydrodynamic model (44) ensures that part of the precipitate, generated in the solute through homogeneous nucleation, is transferred to the mineral surface by adsorption. One should notice that in this model the precipitate $C_{CaCO_{3(p)}}$ is both advected and diffused. Such diffusion enables to account for the interaction of the precipitates with potential unresolved roughness or features in the porous matrix $\Omega_S$.

In the next Section 5, we apply this two-step crystallization model to DNS of $CO_2$ mineral trapping into a real rock geometry at the pore-scale. We investigate the morphological changes in the porous matrix structure, the clogging of pore throats, and the evolution of the macro-scale properties, namely, the porosity and permeability.

## 4.4 High-performance-computing aspects

One of the major constraints when dealing with a semi-Lagrangian formulation lies in the ability of the computational framework to handle an overall hybrid approach in terms of grid-particle formalism, numerical methods, multi-grid resolutions, and hardware devices. Indeed, Cottet et al. (2009) suggested a semi-Lagrangian method coupled with hybrid grid resolutions to address a multi-scale transport problem of a passive scalar. The scalar is discretized on a sub-grid compared to the velocity and vorticity fields and enables the accurate prediction of the small-scale effects. Finally, considering hybrid computing methodologies makes it possible to distribute the distinct parts of an overall problem to different types of hardware architectures. This formalism, therefore, exploits the advantages of each method individually according to the characteristics of the problem.

Nonetheless, implementing such a hybrid approach requires a highly flexible computational framework that gathers a wide range of numerical methods and, therefore, benefits from their intrinsic strengths. One also needs libraries incorporating effective parallel computing tools and able to address, *inter alia*, hybrid CPU-GPU programming. In this section, we present the HPC framework considered to address this DNS of pore-scale reactive flows for $CO_2$ mineral trapping into carbonate rocks.

The library HySoP is a high-performance computing platform (Etancelin et al., 2022), jointly developed at LMAP (Laboratoire de Mathématiques et de leurs Applications, UMR 5142 CNRS, UPPA), LJK (Laboratoire Jean Kuntzmann, Alpes-Grenoble University, UMR 5224 CNRS), and M2N (Laboratoire Modélisation mathématique et numérique, Conservatoire National des Arts et Métiers–CNAM, Paris, EA 7340 CNRS). The library, originally developed to address flow simulations based on remeshed particle methods on hybrid muti-CPU and multi-GPU architectures, was initiated by the work of Etancelin (2014) and has been successfully extended to a larger scope of HPC applications including dissolution at pore-scale (Etancelin et al., 2020).

The entire code is structured around the operator splitting strategy that defines the different operators involved in a problem (at a high level of abstraction), and afterward, enables the discretization of these operators, which are solved using the most appropriate numerical method (at a lower level of abstraction). The overall problem is formally described through an acyclic graph that expresses the operators interactions in the splitting formulation through data dependencies (Etancelin, 2014; Etancelin et al., 2020; Keck, 2019). Even if the code is mainly written in Python, the numerical methods are either implemented using compiled language that enables threads parallelism (using OpenMP on CPU and OpenCL on accelerators) or taken from external libraries. User interface enables building together both numerical methods provided with HySoP on cartesian grid and user defined Python functions. In the latter, performances are obtained provided the usage of interfaces to compiled language (*i.e.*, among others f2py, swig, cython, numba, …). A complementary distributed parallelism is naturally available using domain decomposition implemented with a Message Passing Interface library (MPI).

One of the core features of HySoP is to target hybrid computing using both CPU and GPU. The latter was made possible by the emergence, in the 2000s, of the so-called GPGPU concept, which integrates the GPU as a CPU co-processing partner targeting accelerated performances. The choice of OpenCL was made to cope with portability of the performances as a generic multicore artichecture programming standard (Stone et al., 2010). The initial GPU computing feature was latter enhanced with code generation from templates or symbolic mathematical expressions together with auto-tuning tanks to micro-benchmarks at run time (Keck, 2019). *Just-in-time* compiling is extensively used to achieve performance portability and lazy specification of kernel parameters which known as challenging problem widely dependent on the hardware architecture (Dolbeau et al., 2013).

The previous numerical method have been implemented on a hybrid computing strategy. Darcy-Brinkmann-Stokes equation is solved on CPU architecture using full MPI parallelization technique while all the reactive and transport part is computed using OpenCL on GPUs.

Overall profiling is represented on Figure 6 for two cases of the next section. Results are showing that recent hardware is capable of better performances without any changes in the code. We benefit here from OpenCL code generation and micro-benchmarks on accelerator and usual compiler optimizations for CPU part. Significant variation appears in computational time associated to flow steps because larger $Da_{II}^{crys}$ implies more intense geometrical changes and then reorganization of main flowpath, as described later in Section 5. One can notice the quite large proportion of computational time spend in data management specially for reactive transport part. It corresponds to ghosts layers that handle boundary conditions and whole data transposition in order to get the first varying index identical to the current direction of directional splitted advection and remeshing operator. Ghosts layer widths are quite large due to large CFL numbers handled here. Data management on GPU has been identified as a known bottleneck of the code and would be investigated in future high performance computing engineering efforts.

# 5 Results and discussion on clogging/non-clogging regimes

## 5.1 Pore-scale and reactive setup

The present section focuses on the effects of calcite crystallization on changes in the pore geometry, macro-properties, and flow at the pore scale in the context of $CO_2$ mineral trapping. We consider a pore-scale geometry obtained by microtomography measurements from Sheppard and Prodanovic (2015) and freely available on the Digital Rocks data portal, which includes $\mu CT$ datasets of limestone, glass bead pack, and Castlegate sandstone. The numerical simulations are performed on the Castlegate geometry at a resolution of $128^3$ with a voxel size of 5.6 $\mu$m, which represents a numerical sub-sample of about $L = 0.7168$mm. We assume, as previously introduced in Section 4.2, that the porous matrix is of identical mineral nature as the crystal generated along the reactive process and, thereby, consider that the sub-sample is composed of calcite. While including mineral heterogeneities as perspectives, we here hypothesize the homogeneity of the mineral structure within the sample. Finally, the specific area is numerically estimated for this sample to get, at the initial state, $A_s = 8300$ m$^{-1}$, and is afterward updated along the reactive process.

Numerical simulations are performed under atmospheric conditions in terms of pressure and temperature and rely on the experimental identification of the reaction rate constants, arising from the literature (Chou et al., 1989). We consider isothermal conditions with a temperature of $T = 25°C$ and an injection of $CO_2$ with a partial pressure of $P_{CO_2} = 3.15 \times 10^{-2}$ bar $= 2.96 \times 10^{-2}$ atm — which is about 100 times greater than the partial pressure of $CO_2$ in the atmosphere. Given Henry's law constant for the $CO_2$ at $25°C$, and under the assumption of a highly alkaline medium—with pH > 10.33 — we estimate from Equation (16) that the initial concentration of carbonate ions is given by $C_{CO_3^{2-}}(x, t = 0) = 10^{-3}$ mol.L$^{-1}$.

The calcium initial concentration is subsequently determined based on the equilibrium constant $K_{eq} = 10^{-8.48}$ (Chou et al., 1989; Plummer and Busenberg, 1982) to ensure a far-from-equilibrium precipitation regime given by the supersaturation condition $Q \gg K_{eq}$. We assume that the medium pore space is initially filled with a saturated solution wherein the initial concentration of calcium ions is $C_{Ca^{2+}}(x, t = 0) = 0.1$ mol.L$^{-1}$. Therefore, in our case, the saturation in calcium ions $Ca^{2+}$ initially present in the domain is not a limiting factor of the precipitation reaction, and we consider a continuous calcium injection that maintains the supersaturation constraint. Actually, in order to maintain this supersaturation, we will assume that the concentration in $Ca^{2+}$ remains constant at its initial value. Comparable initial conditions and assumptions have been employed in investigating dissolution experiments by Maes et al. (2022), wherein the sample core was initially flooded with a brine solution that had previously reached equilibrium with supercritical $CO_2$.

Regarding the reaction rate constant for the precipitation, we rely on the experimental identification from Chou et al. (1989) such that:

$$K_{-3} = \frac{K_3}{K_{eq}} = \frac{6.6 \cdot 10^{-7}}{10^{-8.48}} = 199 \text{ mol.m}^{-2}.s^{-1}$$

while adsorption frequencies $K_c$ commonly encountered in the literature range from $10^3$ to $10^8$ s$^{-1}$ (Christoffersen and Christoffersen, 1990; Nielsen, 1984; von Wolff et al., 2021; Wolthers et al., 2012). In practice, we set for the numerical simulations $K_c = 10^3$ s$^{-1}$, the molecular diffusion $D_m = 10^{-9}$ m$^2$.s$^{-1}$ for all the species and the prescribed flow rate $\overline{u} = 1.10^{-3}$ m.s$^{-1}$.

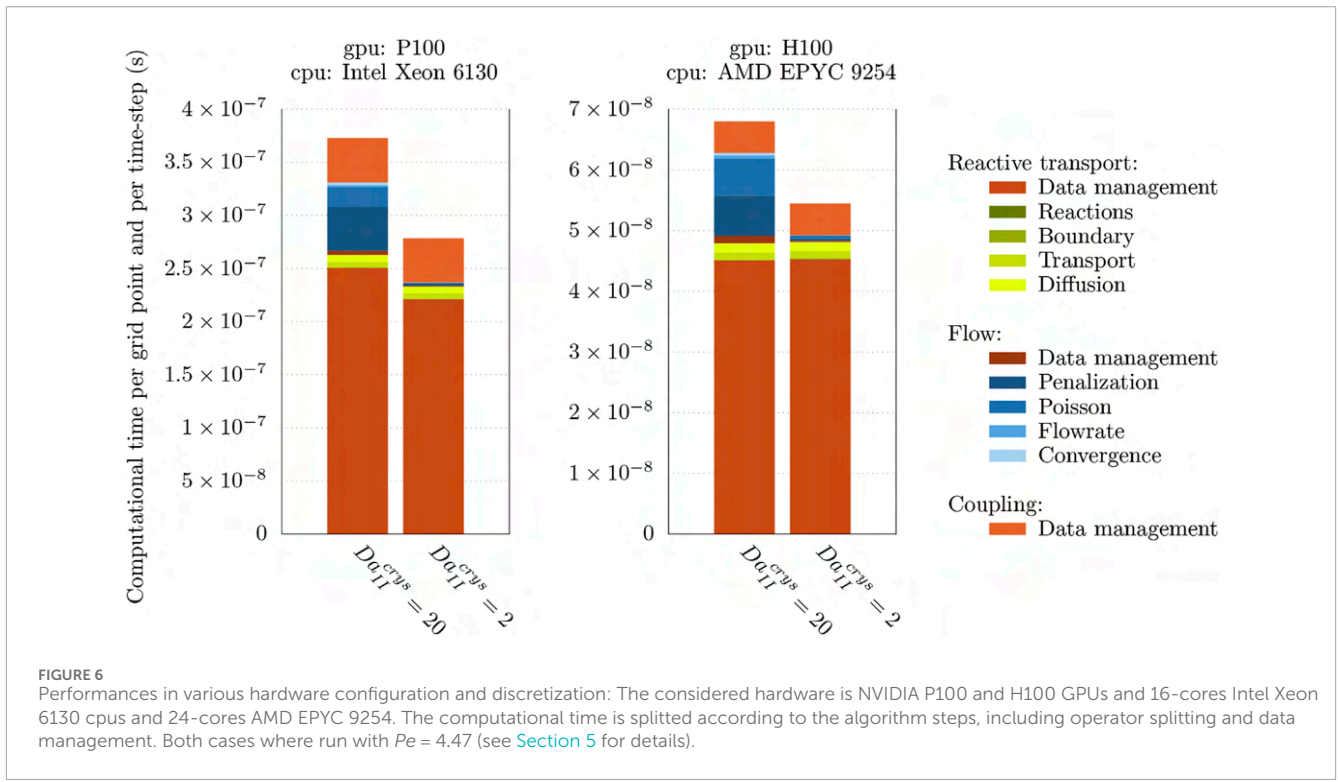

FIGURE 6
Performances in various hardware configuration and discretization: The considered hardware is NVIDIA P100 and H100 GPUs and 16-cores Intel Xeon 6130 cpus and 24-cores AMD EPYC 9254. The computational time is splitted according to the algorithm steps, including operator splitting and data management. Both cases where run with Pe = 4.47 (see Section 5 for details).

## 5.2 Pertinent non-dimensional numbers

Investigating the different precipitation patterns and regimes relies on the definition of well-established characteristic dimensionless numbers, namely, the Peclet and (second or catalytic) Damköhler numbers respectively denoted Pe and $Da_{II}$ (Noiriel et al., 2021; Soulaine et al., 2017; Steefel and Lasaga, 1990). However, in the context of the joint precipitation and crystal growth modeling, we define two distinct Damköhler numbers characterizing each process and respectively denoted $Da_{II}^{prec}$ and $Da_{II}^{crys}$. These dimensionless numbers are subsequently defined as:

$$\text{Pe} = \frac{\bar{u}L}{D_m}, \quad Da_{II}^{prec} = \frac{K_{-3}\gamma_{CO_3^{2-}} A_s L^2}{D_m} \quad \text{and} \quad Da_{II}^{crys} = \frac{K_c L^2}{D_m}$$

where $\bar{u}$ and $L$ are respectively the characteristic velocity and length of the sample, and the activity coefficient of the carbonate ions is $\gamma_{CO_3^{2-}} = 10^{-3}$ m$^3$.mol$^{-1}$. The characteristic length $L$ can be related to pore size (Steefel and Lasaga, 1990), though it is commonly set as $L = \sqrt{\kappa_0}$ provided an experimental or numerical estimation of $\kappa_0$ (Soulaine et al., 2017). The latter alternative is applied here, with an estimation of $\kappa_0 = 2 \times 10^{-11}$ m$^2$ for the porous sample considered.

## 5.3 Evidence of two clogging regimes at same $Da_{II}^{prec}$ and two different $Da_{II}^{crys}$

The first crystallization regime we investigate is characterized by the following dimensionless numbers Pe = 4.47, $Da_{II}^{prec}$ = 33.034 and $Da_{II}^{crys}$ = 20. Precipitation and crystallization of calcite lead to a significant decrease in the macro-scale permeability and porosity, resulting from flow path disruptions at the micro-scale through partial or complete clogging of pore throats. This can also affect the roughness of the mineral interface and the pore-size distribution of the sample and, thereby, contribute to influencing the sample hydrodynamics properties. In particular, we observe these effects at the pore scale in Figure 7 along the reactive process and for several physical times $t$ going from 2h45 to $T_f$ = 6h56. On the right side of Figure 7, we depict partial views of the porous sample's morphology, illustrating the changes in pore structure over the reaction time, along with the micro-porosity field $\varepsilon$ within the porous matrix $\Omega_S$. On the left side, we represent along a slice in the main flow direction (taken at $z = -0.0168$mm), the local variations on the micro-porosity with respect to the initial state—before the reaction process—given by $\varepsilon(t) - \varepsilon(0)$. Initially, we notice that higher micro-porosity variations are more likely localized at the mineral interfaces but also near thin pore throats.

These variations subsequently lead to pore-clogging and reorganization of the main flow pathway (see Figure 7C). In Figure 8B, we investigate the effects of such micro-scale changes on the evolution of the petrophysical properties at the macro-scale, namely, the porosity $\phi$ and permeability $\kappa_0$ (upscaled quantities from Figure 1). The results are consistent with the expected decrease along the $CO_2$ mineral trapping process but also highlight sharp permeability drops, which characterize the pore-clogging phenomena. Finally, in order to identify more clearly the crystallization pattern in this particular regime, we display in Figure 8A an isosurface of the micro-porosity variation between the final and initial times. This illustrates that micro-scale variations occur preferentially in a compact and non-uniform manner in the first inlet part of the domain while following the individual ramifications in the pore structure.

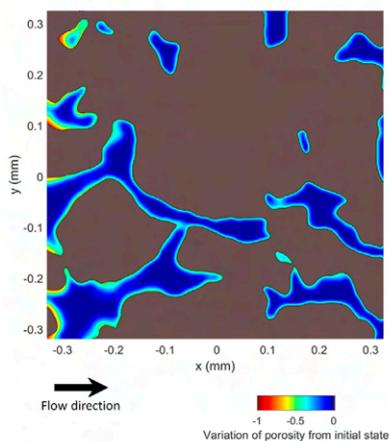
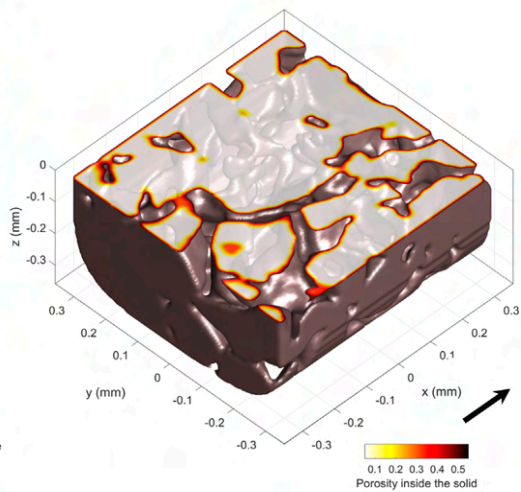
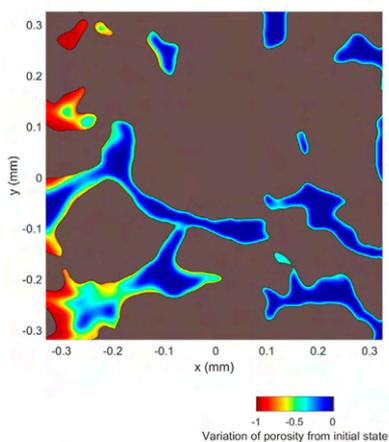
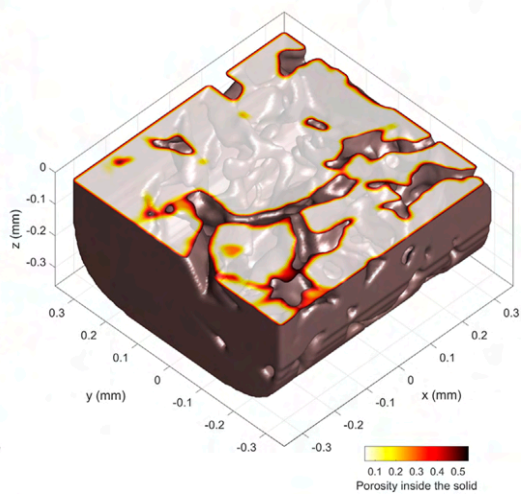
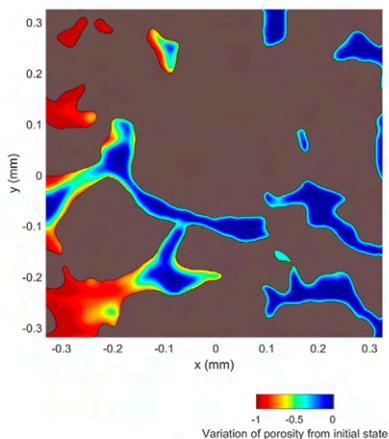
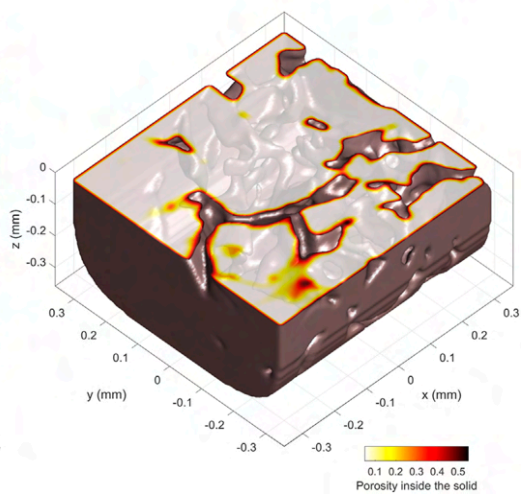

FIGURE 7
Time evolution of the sample geometry and micro-porosity at the pore scale, illustrating pore-clogging effects. Slice at $z = -16.8\mu$m of the porosity variations $\varepsilon(t) - \varepsilon(0)$ in the fluid region of the pore space for various times $t$, on the left. Partial view of the pore space structure as an isosurface of $\varepsilon(t)$ for several times $t$, on the right. The clogging occurs in the red areas, mainly upstream (flow direction is given by the arrow on picture **(A)** Time $t =$ 2h45. **(B)** Time $t =$ 5h30. **(C)** Time $t = T_f =$ 6h56.

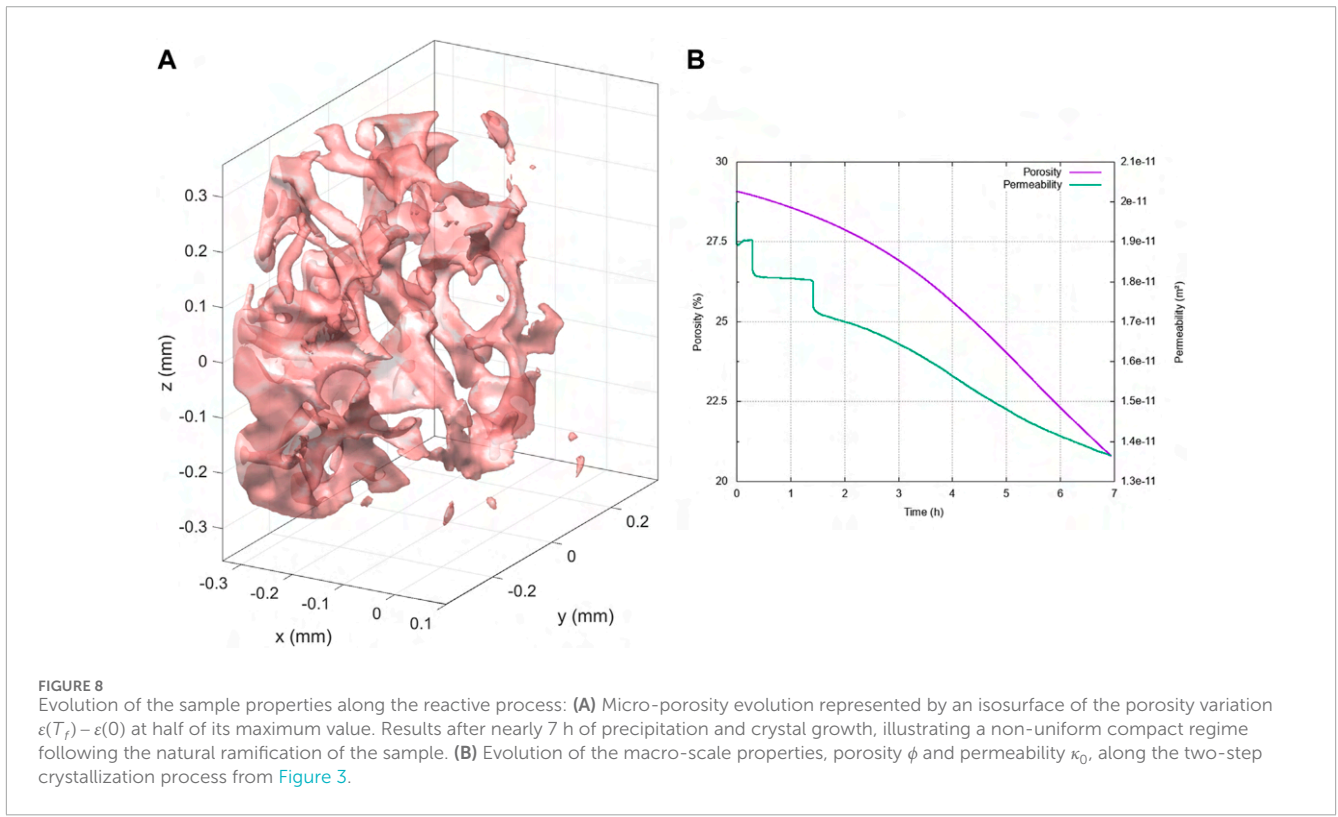

FIGURE 8
Evolution of the sample properties along the reactive process: **(A)** Micro-porosity evolution represented by an isosurface of the porosity variation $\varepsilon(T_f) - \varepsilon(0)$ at half of its maximum value. Results after nearly 7 h of precipitation and crystal growth, illustrating a non-uniform compact regime following the natural ramification of the sample. **(B)** Evolution of the macro-scale properties, porosity $\phi$ and permeability $\kappa_0$, along the two-step crystallization process from Figure 3.

We subsequently investigate the impact of different dominant regimes on the overall two-step crystallization process. To do so, we consider both transport dominant cases with Pe = 4.47 > 1 and diffusion dominant cases with Pe = 0.447 < 1, coupled with two different crystal growth regimes characterized by $Da_{II}^{crys} = 2$ and $Da_{II}^{crys} = 20$. One should notice that the effect of precipitation Damköhler $Da_{II}^{prec}$ changes are not analyzed since this number, characterizing the first homogeneous nucleation regime, is a limiting factor of the crystallization process from Figure 3. Here, we assume in all the previous cases that $Da_{II}^{crys} < Da_{II}^{prec}$ which guarantees a supersaturation state suitable to the development of crystal aggregates on the mineral surface.

To the best of our knowledge, considering that the crystallization regime can be driven by three distinct dimensionless numbers is one of the contributions of the present article. Indeed, most of the regime diagrams presented in the literature mainly characterize precipitation patterns according to the Pe and $Da_{II}^{prec}$ dimensionless numbers, which implies neglecting the effects of nuclei adsorption at the mineral surface in the different regime configurations (Tartakovsky et al., 2007; Yang et al., 2021). However, our results indicate that both homogeneous calcite nucleation (the precipitation step in Figure 3) and growth stages are important in the development of precipitation and crystallization patterns. In particular, we establish that the crystal growth Damköhler number $Da_{II}^{crys}$ has a non-negligible impact on precipitation pattern and porosity variations along the reactive process, regardless of the other dimensionless numbers Pe and $Da_{II}^{prec}$ commonly investigated.

Indeed, Figures 9B, 10B highlight two distinct crystallization regimes at similar Pe and $Da_{II}^{prec}$, but with a ten times smaller adsorption frequency $K_c$ in Figure 9B which impact the $Da_{II}^{crys}$. Figure 9B shows that, for a small adsorption frequency $K_c$, the calcite precipitate is uniformly generated and advected along the main flow path direction (due to the transport dominant regime with Pe > 1) while the micro-porosity changes are minimal. This illustrates that the main flow path is maintained since few calcite nuclei aggregate to the mineral surface. By increasing the adsorption frequency in Figure 10B, the precipitate formation concentrates on the inlet of the domain and is less subject to advection, while the micro-porosity changes become significant. On the one hand, this highlights pore-clogging that prevents further transport of the calcite nuclei. On the other hand, we also notice a backward increase in calcite precipitates that accumulate behind the clogging after some time. The same analysis holds for Figures 9A, 10A, except we consider a diffusion dominant regime with Pe < 1, which means that the precipitate transport is reduced so that the nuclei formation and micro-porosity changes are even more constrained to the inlet boundary of the domain. The results presented in Figures 7 and 8 correspond to the crystallization regime identified by Pe > 1 and $Da_{II}^{crys} = 20$ in Figure 10B. This confirms that the permeability drops observed in Figure 8B are characteristics of pore-clogging effects.

To conclude, at $Da_{II}^{crys} = 2$, $Da_{II}^{prec} = 33$ and for Pe = 0.447 < 1 as well as for Pe = 4.47 > 1, Figure 9 shows that there is no clogging measured. At $Da_{II}^{crys} = 20$, $Da_{II}^{prec} = 33$ and for the two same Peclet numbers, Figure 10 displays evidence of pore-clogging. The clogging is consequently driven principally by the non-dimensional number $Da_{II}^{crys}$ in this case study, the Peclet number affecting the depth of the clogging effect. This is valid as long as the $Da_{II}^{prec}$ is sufficiently high in order to provide precipitate able to turn into crystal.

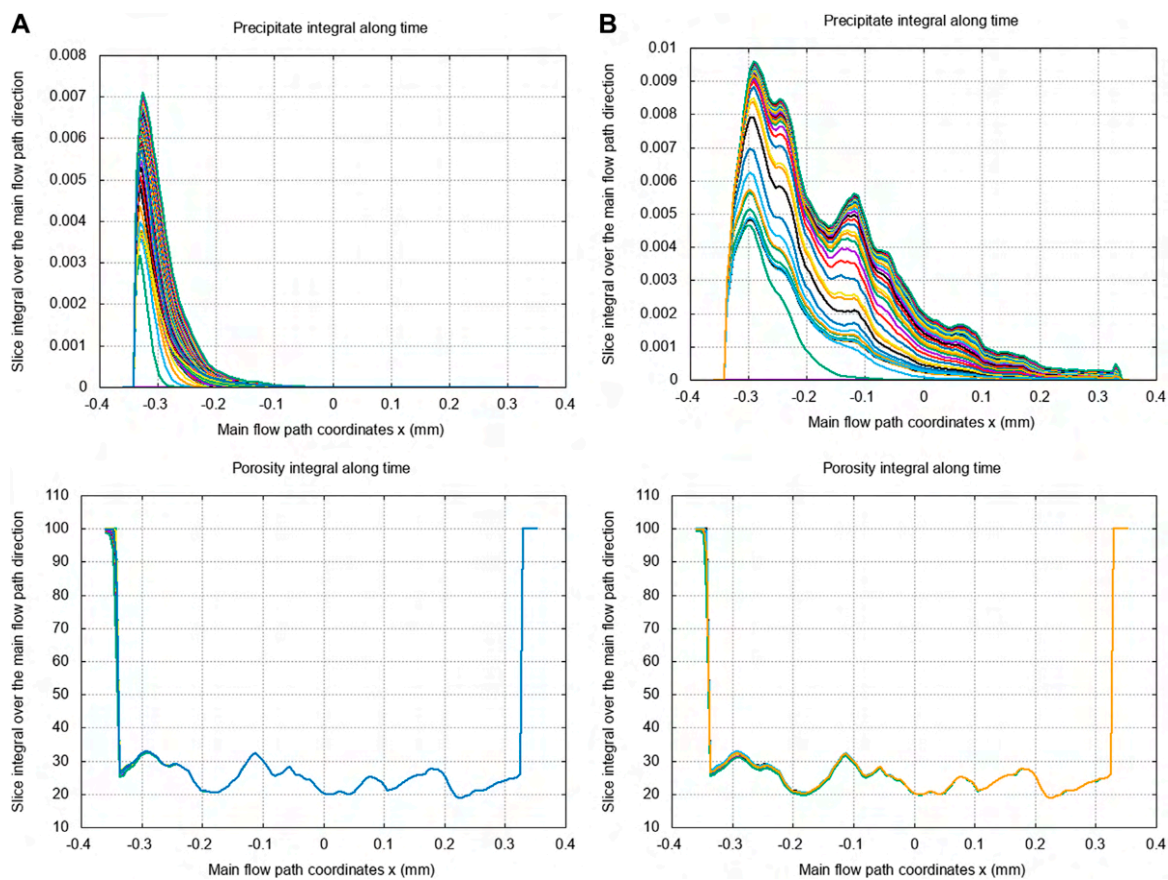

FIGURE 9
Crystallization regimes at $Da_{II}^{crys} = 2$ and $Da_{II}^{prec} = 33$ for Peclet numbers below 1 **(A)** and above 1 **(B)**: no clogging. Slice integrals of the precipitates and macro-porosity—computed over 2D YZ directional slices—plotted with respect to the main flow path direction coordinates x (in mm mm) and where each curve (using segmented colors) represents a distinct time in the reactive process. The porosity integral remains close to its initial value, which shows that pores remain open with the same flow, hence no clogging effect.

## 5.4 Discussion

This section aims at positioning the final set of Equation 21 with the existing literature on nucleation. These models involve the physics of fluid precipitation (homogeneous nucleation) and its subsequent crystallization (heterogeneous nucleation), while as shown on Figure 3, the direct crystallization from reactants is also another possible kind of nucleation. The later direct crystallization is not considered in the present study, but can be compared to our two-step modeling as long as $Da_{II}^{prec} \gg Da_{II}^{crys}$ since we don't model the precipitate rheology.

Moreover, the existing literature considers either internal flows (porous medium as a pore collection) or external flows (porous medium as a grain collection, also referred to as pore-scale). Every study has its own configuration and Dahmköhler/Peclet setup that we cannot investigate one by one in this study.

More in detail concerning recent literature, confined internal flows have been studied experimentally in Poonoosamy et al. (2023); Noiriel et al. (2021) and numerically in Molins and Knabner (2019), but most geometries of porous media described at their pore-scale for systematic analysis are grain-shaped and involve the hydrodynamics of low Reynolds external flow. Such external flow analysis are numerical (Starchenko, 2022; Nooraiepour et al., 2021b; Yang et al., 2021; Varzina et al., 2020; Fazeli et al., 2020) or experimental (Nooraiepour et al., 2021a), some of them focusing on the wall as an initially flat fluid/solid interface Deng et al. (2022). Also, some configuration are in between confined and grain-shaped geometries (Masoudi et al., 2024), with some focus on cement material (Tong et al., 2024; Patel et al., 2021).

Among these studies, qualitative comparison with Yang et al. (2021), and quantitative comparison with Masoudi et al. (2024) are of direct interest for the present investigation.

First, in Yang et al. (2021), a diagram showing the expected nucleation regime with respect to $Da_{II}$ and $Pe$ is available on their Figure 13. They consider crystal growth around a grain, studying the hydrodynamic and reaction balance and the resulting grain shape. Their model rely on Kozeny-Carman correlation law and TST so only the numerical method significantly differs (they use LBM while we use semi-Lagrangian with PSE method). Their results are compatible to ours, as for $1 \leq Da_{II} \leq 100$ and $Pe \leq 10$ we don't observe dendrite formation nor uniform distribution of crystallization.

Second, the article Masoudi et al. (2024) considers only heterogeneous nucleation with both spherical grains and pore structures at several crystallites levels and links porosity and permeability during crystallization process. They consider Lattice Boltzmann simulations of crystal growth from aqueous material with a rate $R_G = k_G S$ in mol/s which is no more than our relation (41). Despite the formulation of crystallites events fitting well to LBM expression of reaction formula but far from our deterministic PDE formulation, we are comparable to their heterogeneous geometry with high crystallite level. Our configuration at $Da_{II}^{crys}$ = 20 contains region of high crystallization rate and some not reached by reactants so the comparison is pertinent to the cases B and D of the Figure 3 from Masoudi et al. (2024). On Figure 11 we plot the similar petro-physical diagram showing the renormalized permeability $K/K_0$ with respect to renormalized global porosity $\phi/\phi_0$. In order to get a pertinent measurement of the crystallization process, we restrict the computation of $K$ and $\phi$ to the upstream quarter of the computational domain, so that the porosity displayed on Figure 10 for Pe > 1 evolves from 30% to 13.9% in 3h40, that is to say $\phi/\phi_0$ = 46.5% of the initial porosity remains.

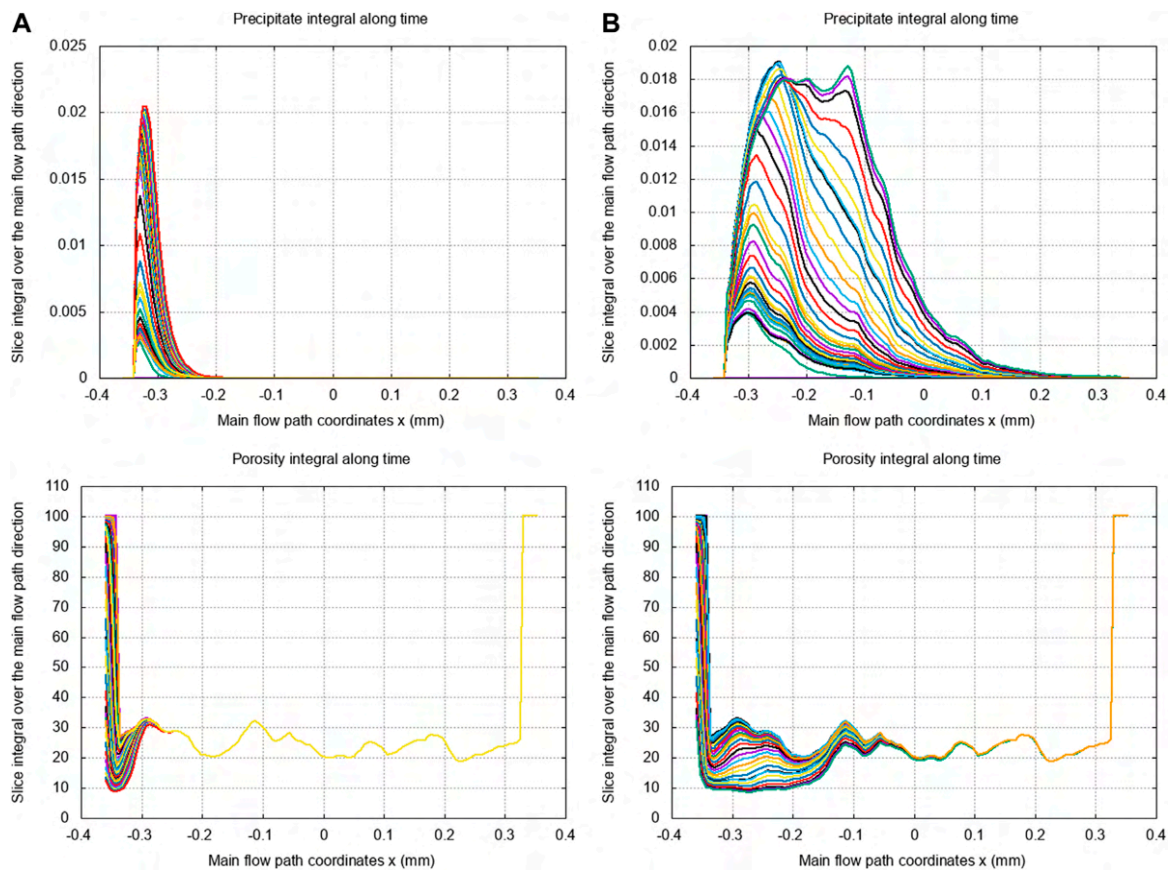

FIGURE 10
Crystallization regimes at $Da_{II}^{crys}$ = 20 and $Da_{II}^{prec}$ = 33 for Peclet numbers below 1 **(A)** and above 1 **(B)**: pore-clogging. Slice integrals of the precipitates and macro-porosity—computed over 2D YZ directional slices—plotted with respect to the main flow path direction coordinates x (in mm mm) and where each curve represents a distinct time in the reactive process. The porosity integral goes from 30% to 10% (close to the cristal porosity at 5%) in the upstream part of the sample, which shows a pore-clogging effect.

With this setup, the renormalized $K - \phi$ diagram is displayed on the Figure 11A, along with the power laws $K/K_0 = (\phi/\phi_0)^n$ for $n = 2, 3, 8, 64$, the case $n = 3$ being related to the Kozeny-Carman correlation far from $\phi = 1$. We exhibits on this diagram our 4 configurations (two different Dahmköhler and Peclet number) for the whole simulation and for the strongest crystallization events, corresponding to individual pore clogging time window and identified by a sharp decrease of the permeability (see Figure 8 for instance). At a global level we observe clogging events with exponents $n$ between 64 and 8 as expected [see Figure 3D in Masoudi et al. (2024)], but a lower exponent around 1.5 at long time scale [2–4 was expected from Figure 3B in Masoudi et al. (2024)]. Concerning the time evolution of our regime, the Figure 11B shows a zoom at high $K$ and $\phi$ values, exhibiting short time scale (the maximum renormalized values 1 corresponding to the initial condition). We can then read that for short time scale at $Da_{II}^{crys}$ = 20, the Kozeny-Carman law is quite well followed but deviates strongly after the last individual pore clogging ($t = 1h25$). Moreover, the low Dahmköhler case $Da_{II}^{crys}$ = 2 that exhibits no clogging (see Figure 9 for instance) has very limited effect on permeability and porosity (see Figure 11B), but shows nevertheless a small crystallization event at low Pecler number that is also coherent in term of exponent.

Moreover, to conclude with this comparison, such a $K - \phi$ criterion can be applied only when the full sample is involved in the crystallization process, otherwise the porosity will reach

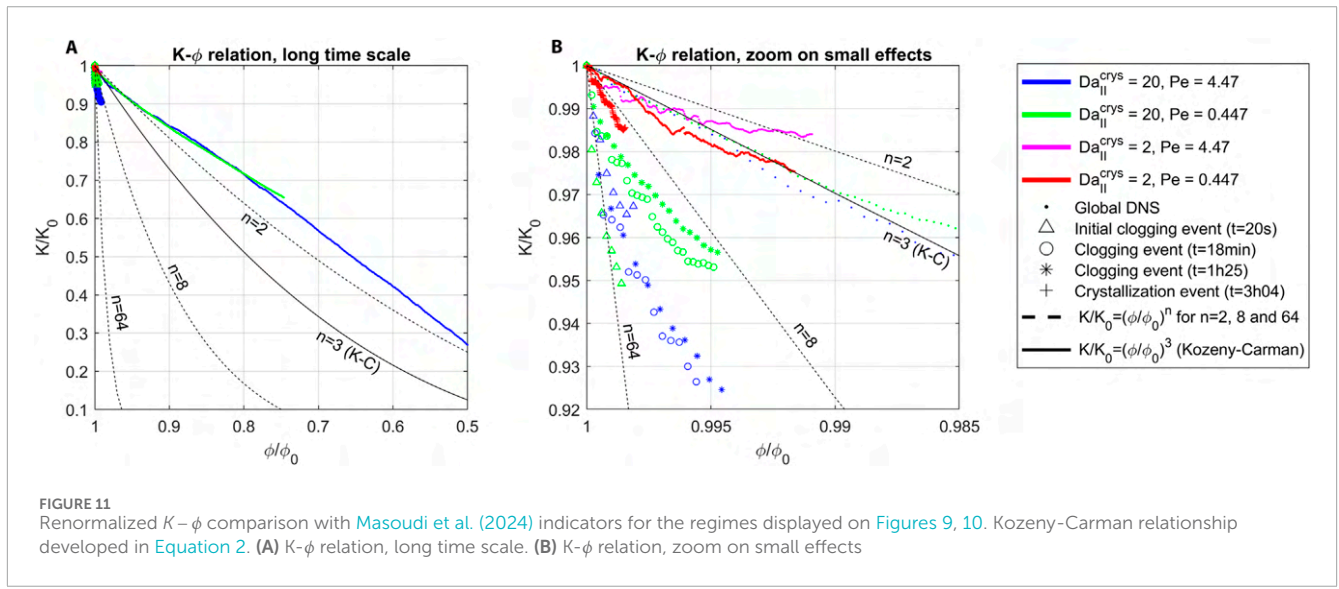

FIGURE 11
Renormalized $K – \phi$ comparison with Masoudi et al. (2024) indicators for the regimes displayed on Figures 9, 10. Kozeny-Carman relationship developed in Equation 2. **(A)** $K-\phi$ relation, long time scale. **(B)** $K-\phi$ relation, zoom on small effects a limit and the permeability will still decrease, which is not compatible with a power-law. Consequently, a special care has to be taken for the computational domain of this criterion when considering a preferential spot of cristallization (such as our case of upstream furring).

Finally, in Starchenko (2022), nucleations occur around a single grain but can be compared to our model (Equation 21) if reduced to 1D (our $Da_{II}^{prec}$ values are similar, 34 and 33 respectively). Deng et al. (2022) shows growth on a wall, also comparable if reduced to 1D. These two growth rates at wall scale will be the topic of future investigation, but are currently too far, at a geometrical level, from our pore-scale configuration.

# 6 Concluding remarks

This article focused on developing an efficient DNS framework to address reactive flows at the pore scale in the context of $CO_2$ mineral storage. Indeed, the injected $CO_2$ will interact with the aquifer structure and eventually lead to mineral trapping in the form of calcite precipitates and crystals. These processes are interesting to study at the pore scale to ensure a comprehensive analysis of the local rock-fluid interactions and evolving pore structures. This can subsequently translate into meaningful estimations of the macro-scale properties changes and measure the impact of the geochemical processes on the natural underground reservoirs. In particular, precipitation and crystallization lead to a significant reduction in the macro-scale permeability and porosity, which result from partial or complete pore clogging and thus from a reorganization of the flow path at the micro-scale.

From a conceptual perspective, we developed a new crystallization model that efficiently combines a classical deterministic TST approach of the nucleation process with a probabilistic view of the crystal aggregation to the pore surface. This enables us to account for spatial and geometrical dependency in the crystal growth modeling through a probabilistic attachment rate depending on local mineral volume fraction. In this sense, we integrate the modeling of preferential growing sites that largely hinge on the surrounding pore arrangement. To the best of our knowledge, such considerations are here accounted for the first time to model crystallization processes in complex 3D geometries at the pore scale. Investigating probabilistic attachment rates based on the surrounding pore structure also ensures reliable prediction of pore-clogging at the pore scale. Finally, we demonstrate that the proper characterization of crystallization regimes both depends on the nucleation process and crystal aggregation. Indeed, we exhibit that the two commonly considered dimensionless numbers, Pe and $Da_{II}^{prec}$, are not sufficient to explain clogging effects and precipitation patterns. A novelty of the present manuscript is, therefore, that the crystallization regimes are characterized by three dimensionless numbers that include the effects of nuclei adsorption to the pore surface.

This reactive hydrodynamic model consistently couples a Lagrangian formulation for the reaction equations with a grid-based approach for the flow using the DBS equation with the superficial velocity formalism. This semi-Lagrangian method is addressed through a splitting operator strategy coupled with high-order remeshing steps for grid-particle interpolations. This original numerical method introduced to solve this coupled model has been efficiently incorporated into the hybrid numerical framework HySoP and results in a CPU-GPU implementation of the method. It includes a robust estimation of the heterogeneous diffusion operator arising from Archie's law term in the reactive system.

At the same time, this article also demonstrated strong implications in the overall reactive system of several parameters that can be subject to a wide range of discrepancies. In particular, morphological features and kinetic parameters, such as the micro-porosity $\varepsilon$, specific area $A_s$, rate constants $K_i$, and adsorption frequencies $K_c$, have a significant impact on the reaction rates and the dynamical patterns.

Experimental determination of these parameters can range over several orders of magnitude and result in highly different regimes that drastically affect the estimation of the macro-scale properties: the adsorption frequencies $K_c$ commonly found in the literature can range from $10^3$ to $10^8$ s$^{-1}$ as shown in Christoffersen and Christoffersen (1990); Nielsen (1984); von Wolff et al. (2021); Wolthers et al. (2012). Our present method has been shown to compute accurately the phenomena of crystallization and precipitation in different regimes, and will be used intensively in future works for inverse problems in order to get a robust and accurate estimation of such adsorption frequencies from experiments imaging. New methods relying of AI and BPINNs for the estimation of reaction rates at the pore scale are already available for dissolution (Perez and Poncet, 2024) and could be adapted to nucleation process in the near future.

## Data availability statement

The raw data supporting the conclusions of this article will be made available by the authors, without undue reservation.

## Author contributions

SP: Conceptualization, Formal Analysis, Investigation, Methodology, Resources, Software, Validation, Writing–original draft, Writing–review and editing. J-ME: Investigation, Resources, Software, Validation, Visualization, Writing–original draft, Writing–review and editing. PP: Conceptualization, Funding acquisition, Investigation, Methodology, Project administration, Supervision, Visualization, Writing–original draft, Writing–review and editing.

## Funding

The author(s) declare that financial support was received for the research, authorship, and/or publication of this article. This work was partially supported by ANR Grant ANR-20-CE45-0022, Carnot Institute ISIFoR Grant P450902ISI and E2S-UPPA project MicroMineral.

## Conflict of interest

The authors declare that the research was conducted in the absence of any commercial or financial relationships that could be construed as a potential conflict of interest.

## Publisher's note

All claims expressed in this article are solely those of the authors and do not necessarily represent those of their affiliated organizations, or those of the publisher, the editors and the reviewers. Any product that may be evaluated in this article, or claim that may be made by its manufacturer, is not guaranteed or endorsed by the publisher.

## Supplementary material

The Supplementary Material for this article can be found online at: https://www.frontiersin.org/articles/10.3389/feart.2025.1493305/full#supplementary-material

CORRESPONDANCE
Philippe Poncet,
philippe.poncet@univ-pau.fr

# A semi-Lagrangian method for the direct numerical simulation of crystallization and precipitation at the pore scale

## Supplementary material: Original and discretization corrected particle-strength-exchange methods

Sarah Perez [1,2], Jean-Matthieu Etancelin [1] and Philippe Poncet [1,*]

[1] Universite de Pau et des Pays de l'Adour, E2S UPPA, CNRS, LMAP, Pau, France, [2] The Lyell Centre, Heriot-Watt University, Edinburgh, United Kingdom

## 1 Introduction

This supplementary section concerns the theoretical aspects of the PSE method that finds its essence in estimating diffusion in a Lagrangian context with mesh-less and scattered particle structures. Since the original article from Degond and Mas-Gallic (1989), the PSE approach has appeared as an efficient numerical method for solving convection-diffusion problems with particles (Schrader et al., 2012) and has been successfully used in vortex methods (Cottet et al., 2000).

Several reviews have also extended its application to the Eulerian context, with structured grids, but also to hybrid grid-particles formalism while enhancing the accuracy of the original method by replacing continuous integration with discrete one, such as in Bergdorf et al. (2005); Poncet (2006); Schrader et al. (2010). These novel PSE approaches, therefore, enable us to efficiently evaluate the heterogeneous diffusion operator arising from Archie's law in a Semi-Lagrangian context. In the following, we briefly review the general principles of this original method along with an overview of its successive improvements using discretization corrections.

We describe in this supplementary section how to build the quadrature of the diffusion operator $\operatorname{div}(\mathbf{L}\nabla f)$ in $\mathbb{R}^d$, with $\mathbf{L}$ a space-variable positive symmetric matrix, under the formulation

$$Q^\xi \cdot f^h(x_k) = \sum_{x_l \in \mathcal{S}(x_k)} \sigma^\xi(x_k, x_l) \left( f_l - f_k \right) v_l \qquad (1)$$

for a measure function (the sum of $N_p$ Dirac functions)

$$f^h = \sum_{i=1}^{N_p} f_i \, v_i \, \delta_{x_i}$$

with $f_i = f(x_i)$ as defined in the main text of the article, and where $\mathcal{S}(x_k) := \operatorname{Supp}\left(\sigma^\xi(x_k, \cdot)\right)$ refers to the set of points in the support of the kernel function $\sigma^\xi$. The two following sections aim at building this exchange kernel $\sigma^\xi$.

## 2 Original particle-strength-exchange method

The PSE scheme is then completely determined once the kernel $\sigma^\xi$ is defined and exhibits its relation with the diffusion matrix $\mathbf{L}$. The original approach from Degond and Mas-Gallic (1989) suggests the following kernel choice

$$\sigma^\xi(x,y) = \frac{1}{\xi^2} \sum_{i,j=1}^{d} M_{ij}(x,y)\psi_{ij}^\xi(y-x),$$

where

$$\psi_{ij}^\xi(x) = \frac{1}{\xi^d}\psi_{ij}\left(\frac{x}{\xi}\right)$$

is a matrix cutoff function with $\psi_{ij}$ symmetric and even, and $\mathbf{M} = (M_{ij}(x,y))$ a symmetric matrix to be determined. These hypotheses are of great interest as they guarantee the conservation property of the operator $Q^\xi$ based on symmetric exchanges. We then introduce the matrix $\mathbf{m}(x) := \mathbf{M}(x,x)$ and the moments of the cutoff functions $\psi_{ij}$ given by:

$$Z_{ij}^\alpha = \int \psi_{ij}(x)x^\alpha dx,$$

for any $i,j$ and multi-index $\alpha$. It has been proved by Degond and Mas-Gallic that if some moment conditions are satisfied, namely the hypotheses (i) and (ii), we obtain the following convergence result (Degond and Mas-Gallic, 1989):

PROPERTY 1. *We assume that there exists an integer $r \geq 2$ such that :*
*(i) $Z_{ij}^\alpha = 0$ for $1 \leqslant |\alpha| \leqslant r+1$ and $|\alpha| \neq 2$*
*(ii) for any integer k,l in $[1,d]$, we have*

$$\sum_{i,j=1}^{d} m_{ij}(x)Z_{ij}^{e_k+e_l} = 2L_{kl}(x)$$

*In addition of the previous hypotheses on matrices $\mathbf{M}$, $\mathbf{m}$ and $\psi$, we assume the following regularities $\mathbf{M} \in W^{r+1,\infty}(\mathbb{R}^d \times \mathbb{R}^d)$, $\mathbf{m} \in W^{r+1,\infty}(\mathbb{R}^d)$ and $(1+|x|^{r+2})\psi(x) \in L^1(\mathbb{R}^d)$. There exists a positive constant $C = C(\mathbf{M},\psi)$ such that for any function $f \in W^{r+2,\infty}(\mathbb{R}^d)$*

$$\|\text{div}(L\nabla f) - Q^\xi \cdot f\|_{0,\infty} \leqslant C\xi^r \|f\|_{r+2,\infty}.$$

Several matrix cutoff functions have been investigated in Degond and Mas-Gallic (1989) but we mainly focus on the most suitable for practical use, which reads as:

$$\psi_{ij} = x_i x_j \Theta(x)$$

with $\Theta$ a smooth spherically symmetric function with fast decreasing, also called the stencil generator. Finally, one needs to define the second moments' matrix of $\Theta$, denoted $\mathbf{A} = (a_{kl})$, and given by

$$a_{kl} = \int x_k^2 x_l^2 \Theta(x)dx, \qquad k,l \in [1,d].$$

In this case, one gets the existence of a matrix $\mathbf{m}(x)$ such that the hypotheses (i) and (ii) of Property 1 are satisfied if and only if $\mathbf{A}$ is an invertible symmetric matrix and, for any k,l $k \neq l$, we have $a_{k,l} \neq 0$. The matrix $\mathbf{m}(x)$ is then defined by (see Lemma 1 in Degond and Mas-Gallic (1989)):

$$m_{kl}(x) = a_{kl}^{-1} L_{kl}(x), \quad \text{for} \quad k \neq l \in [1,d] \quad (2)$$

$$\sum_{i=1}^{d} a_{ki}m_{ii}(x) = 2L_{kk}(x), \quad \text{for} \quad k \in [1,d] \quad (3)$$

which is a fundamental result of the original PSE article. In 3D applications, for instance, one can compute the matrix $\mathbf{A}$ coefficients using spherical coordinates to obtain $a_{kk} = 3\gamma$ and $a_{kl} = \gamma$, if $k \neq l$, with $\gamma$ expressed by:

$$\gamma = \frac{4\pi}{15} \int_0^\infty \widetilde{\Theta}(r)r^6 dr$$

where the spherically symmetric function $\Theta$ is written $\Theta(x) = \widetilde{\Theta}(|x|)$. Solving the problem given by equations (2) and (3) then explicitly provides

$$m_{kl} = \gamma^{-1}L_{kl}, \quad \text{if } k \neq l,$$

and

$$m_{kk} = \gamma^{-1}L_{kk} - \frac{\gamma^{-1}}{5}Tr(\mathbf{L}),$$

which also writes

$$\mathbf{m} = \gamma^{-1}\mathbf{L} - \frac{\gamma^{-1}}{5}Tr(\mathbf{L})\mathbf{Id}_3. \quad (4)$$

When the conditions (2) and (3) are satisfied and in the case of a kernel defined with a spherical-symmetric function $\Theta$ such as in (2), the method provides at least second order approximation of the diffusion operator (see Property 1) which is suitable for any particle distribution in a Lagrangian context. One should notice that in practice the method is limited to second-order convergence for positive kernel $\sigma^\xi$ (Cortez, 1997; Degond and Mas-Gallic, 1989).

Finally, to make the approximation operational, it remains to define the relation between $\mathbf{M}$, $\mathbf{m}$, $\mathbf{L}$, and $\Theta$. A usual approach is to consider the matrix $\mathbf{M}(x,y) = (\mathbf{m}(x) + \mathbf{m}(y))/2$ and for $\psi$ given by (2), one gets the following kernel formula:

$$\sigma^\xi(x,y) = \frac{1}{\xi^{d+4}}\Theta\left(\frac{y-x}{\xi}\right)\mathbf{M}(x,y) : (x-y)^{\otimes 2} \quad (5)$$

This entirely specifies the numerical scheme by combining the equations (1), (5), and (4). This original formalism is based on continuous integration of the second moments of $\Theta$ through the equation (2) and is second order consistent in the sense $\mathcal{O}[(h/\xi)^2]$.

## 3 Discretization corrected particle-strength-exchange (DC-PSE) method

Alternatives relaxing this constraint on convergence result in replacing the continuous moment integration with discrete moment conditions, which is referred to as discretization correction of the PSE scheme (DC-PSE). The latter has been successfully developed for several state-of-the-art applications including mesh-free scenarios with arbitrary particle distributions, uniform Cartesian grids, and also in hybrid formulations involving both an underlying grid along with the particles, detailed in Bergdorf et al. (2005); Bourantas et al. (2016); Schrader et al. (2010); Zwick et al. (2023).

The consistency of the original PSE scheme can be improved to $\mathcal{O}(h^2)$ by using discrete integration when particles are distributed over a uniform grid (Poncet, 2006). The main idea is to replace (2) by the matrix of discrete second-order moments, which subsequently leads to a distinct evaluation of the matrix $\mathbf{m}$. One introduces the following coefficients using discrete integration:

$$\gamma_1 = \sum_{x \in \mathbb{J}} x_k^4 \Theta(x) h^d, \quad k \in [1, d]$$

$$\gamma_2 = \sum_{x \in \mathbb{J}} x_k^2 x_l^2 \Theta(x) h^d, \quad k \neq l \in [1, d]$$

for $\mathbb{J} \subset h\mathbb{Z}^d$ a $d$-dimensional lattice, including at least one neighborhood of the current mesh point. In the main text of the article, $d = 3$ is used for these formula. One then gets similar equations as (2) and (3) with respect to the coefficients $\gamma_1$ and $\gamma_2$, resulting in the characterization of $\mathbf{m}$. We introduce the matrix $\mathbf{H} = H_{ij}$ given by

$$H_{ij} = \left( \frac{\gamma_1^2 - \gamma_1 \gamma_2 - 6\gamma_2^2}{\gamma_2(\gamma_1^2 + \gamma_1 \gamma_2 + 2\gamma_2^2)} \right) (1 - \delta_{ij}) L_{ij}$$

where $\delta_{ij}$ is the Kronecker symbol, such that the matrix $\mathbf{H}$ is zero when $\mathbf{L}$ is diagonal or when $\gamma_1 = 3\gamma_2$. We thus obtain a discrete renormalization of the matrix $\mathbf{m}$ which reads as follows:

$$\mathbf{m} = c_0 \mathbf{L} - c_1 \mathrm{Tr}(\mathbf{L}) \mathbf{Id_3} + \mathbf{H}$$

where

$$c_0 = \frac{2(\gamma_1 + 2\gamma_2)}{\gamma_1^2 + \gamma_1 \gamma_2 - 2\gamma_2^2} \quad \text{and} \quad c_1 = \frac{2\gamma_2}{\gamma_1^2 + \gamma_1 \gamma_2 - 2\gamma_2^2},$$

and replaces the equation (4) in the original PSE version. Finally, this formulation with discrete integration is completely defined through the formula (5). This also leads to a better accuracy since this scheme is consistent in $h^2$ whereas the classical PSE method has a convergence in the sense that the error is of order $(h/\xi)^2$.

While the discrete renormalization of PSE method can be assimilated to a FD stencil on a uniform Cartesian grid and satisfies the same order of accuracy as standard FD schemes, one can query the motivation for using this seemingly complex approach.

To summarize, behind appearances, this method is easy to implement and not computationally expensive in the context of uniformly distributed grids. In 3D applications, for instance, the PSE discrete formulation (1) can be finally written as follows

$$Q^\xi \cdot f^h(x_k) = \frac{1}{\xi^7} \sum_{l \sim k} (f_l - f_k) \Theta\left(\frac{x_l - x_k}{\xi}\right) \times \left[ \sum_{i,j=1}^{3} M_{ij}(x_k, x_l)(x_l - x_k)_i (x_l - x_k)_j \right] v_l$$
(6)

with the spherically symmetric function $\Theta(x) = 1/(1 + |x|^p)$ where $|\,.\,|$ is the Euclidean norm in $\mathbb{R}^3$.

The formula (6) basically involves all the contributions of the mesh points of index $l$ in the $\xi$-neighborhood of the current mesh point $x_k$, representing namely 26 neighbors in 3D for $\xi = h$ compared to merely 6 neighbors with standard crossed FD scheme. In practice, $\xi$ is taken equal to $h$ or $2h$, and the numerical results presented in the next section are given with $p = 10$.

## 4 Validation of the PSE scheme for heterogeneous diffusion

In order to validate the formulation and the implementation, we consider the approximation of a diffusion operator $\mathrm{div}(\mathbf{L}\nabla f)$ on a domain $\Omega \subset \mathbb{R}^3$ with a spatially varying diffusion matrix

$$\mathbf{L}(x, y, z) = (1 + \cos^2(x)) \mathbf{I}_3$$

and a function $f$ given by $f(x, y, z) = \sin(x) \sin(y) \sin(z)$. This ensures three-periodic boundary conditions on the domain $\Omega = ]-\pi; \pi[^3$. Direct computation of the exact solution provides

$$\mathrm{div}(\mathbf{L}\nabla f)(x, y, z) = -3 f(x, y, z) - 5\cos^2(x) f(x, y, z)$$

and a mesh convergence analysis is performed, studying the error norm against the mesh step $h$, for mesh resolutions going from $32^3$ to $256^3$. Introducing the discrete error vector $\mathrm{Err}$, defined for each point $X_k \in \mathbb{R}^3$ of the

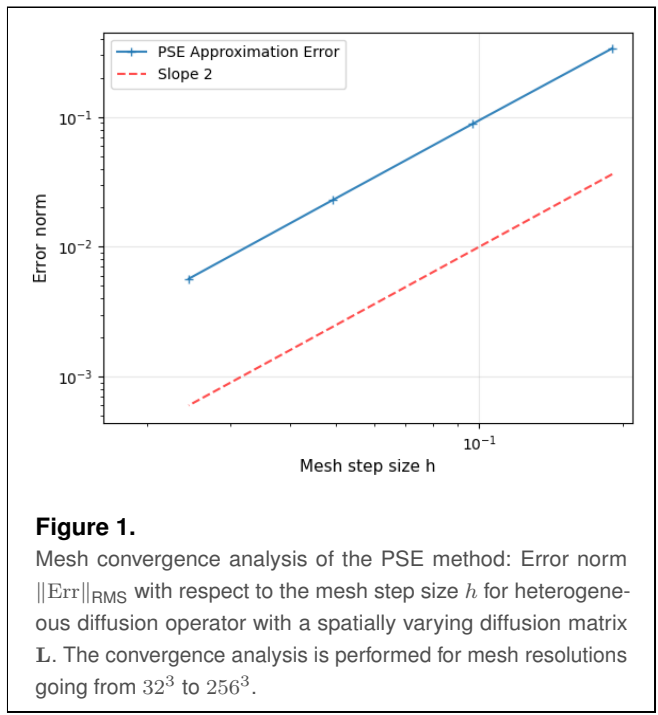

**Figure 1.**
Mesh convergence analysis of the PSE method: Error norm $\|\mathrm{Err}\|_{\mathrm{RMS}}$ with respect to the mesh step size $h$ for heterogeneous diffusion operator with a spatially varying diffusion matrix **L**. The convergence analysis is performed for mesh resolutions going from $32^3$ to $256^3$.

grid by

$$\mathrm{Err}_k := \mathrm{div}(\mathbf{L}\nabla f)(X_k) - Q^\xi \cdot f^h(X_k),$$

we compute for each mesh resolution the RMS norm $\|\mathrm{Err}\|_{\mathrm{RMS}}$ inherited from the functional $\mathbb{L}^2$-norm on $\Omega$. The value of $Q^\xi \cdot f^h$ is defined on every particles by formulae (1) or (6), despite the fact that $f^h$ is a measure function, that is to say a combination of Dirac functions. We retrieve a second-order convergence curve presented in Figure 1.